\documentclass[a4paper,11pt]{article}

\usepackage[T1]{fontenc} 
\usepackage[utf8]{inputenc}
\usepackage{jheppub}
\usepackage{amsmath}
\usepackage{amssymb}
\usepackage{enumitem}
\usepackage{graphicx}
\usepackage{subcaption}
\usepackage{color}

\usepackage{dcolumn}
\usepackage{bm}
\usepackage{cleveref}
\usepackage{epsfig}
\usepackage{yfonts}
\usepackage{color}
\usepackage{soul}
\usepackage[justification=centering]{caption}
\usepackage{mathrsfs} 

\newcommand{\be}{\begin{equation}}
\newcommand{\ee}{\end{equation}}

\makeatletter
\renewcommand*\env@matrix[1][\arraystretch]{%
	\edef\arraystretch{#1}%
	\hskip -\arraycolsep
	\let\@ifnextchar\new@ifnextchar
	\array{*\c@MaxMatrixCols c}}
\makeatother

\title{Non-equilibrium steady states in quantum critical systems with Lifshitz scaling}

\author[a]{Daniel Fern\'andez,}
\author[a]{Aruna Rajagopal,}
\author[a,b]{L\'arus Thorlacius}
\affiliation[a]{University of Iceland, Science Institute, Dunhaga 3, 107 Reykjav\'ik, Iceland}
\affiliation[b]{The Oskar Klein Centre for Cosmoparticle Physics \& Department of Physics,\\ Stockholm University, 
	AlbaNova, 106 91 Stockholm, Sweden.}
\emailAdd{arr17@hi.is}
\emailAdd{fernandez@hi.is}
\emailAdd{lth@hi.is}

\abstract{We study out-of-equilibrium energy transport in a quantum critical fluid with Lifshitz scaling symmetry 
following a local quench between two semi-infinite fluid reservoirs. The late time energy flow is universal and is 
accommodated via a steady state occupying an expanding central region between outgoing shock and rarefaction 
waves. We consider the admissibility and entropy conditions for the formation of such a non-equilibrium steady 
state for a general dynamical critical exponent $z$ in arbitrary dimensions and solve the associated Riemann problem. 
The Lifshitz fluid with $z=2$ can be obtained from a Galilean boost invariant field theory and the non-equilibrium steady 
state is identified as a boosted thermal state. A Lifshitz fluid with generic $z$ is scale invariant but without boost 
symmetry and in this case the non-equilibrium steady state is genuinely non-thermal.
}

\begin{document}

\maketitle
\flushbottom



\section{Introduction}

Fluid theory is one of the oldest effective descriptions in physics.\footnote{For a classic textbook treatment see volume~6 
	of Landau and Lifshitz \cite{landau_lifshitz_6}.} It is based on general symmetry principles and applies in the limit of long 
wavelength and low frequency compared to characteristic microscopic length and time scales of the system in question.   
A fluid description can thus stand on its own and be useful even when no microscopic description, based on particles or 
quasiparticles, is available. There has been considerable recent interest in extending fluid theory to systems with 
unconventional symmetries, including Lifshitz scale symmetry, with potential applications to quantum 
critical systems \cite{Hoyos:2013eza,Hoyos:2013qna,Kiritsis:2015doa,Hartong:2016nyx}. 
Motivated by these developments, we will consider a problem involving out of equilibrium 
energy transport in fluids with Lifshitz symmetry. 

It remains an open problem to develop a general fluid dynamics formalism for systems that are far from thermal equilibrium, 
but there has been interesting recent progress in this direction involving relativistic fluids. Investigating out of equilibrium 
energy transport between two relativistic quantum critical heat baths led to the discovery of the emergence of a universal 
Non-Equilibrium Steady State (NESS) between the two heat baths, described by a Lorentz boosted thermal 
state \cite{Bernard:2012je,Bhaseen:2013ypa,PhysRevD.94.025004,Spillane:2015daa,Pourhasan2016}. 
In the present paper, we extend this analysis to more general quantum critical fluids, in particular to non-relativistic 
fluids with Lifshitz scale symmetry (referred to as Lifshitz fluids in the following), and find that a NESS emerges here as well. 
For the special case of a Lifshitz fluid with dynamical critical exponent $z=2$, the resulting NESS can be viewed as
a Galilean boost of a thermal state. For Lifshitz fluids with $z\neq 2$, there is no underlying boost 
symmetry \cite{10.21468/SciPostPhys.5.1.003}. It turns out there is still an emergent NESS at generic $z$, 
but in this case it cannot be obtained as a boosted thermal state. 

In order to gain further insight into emergent hydrodynamic behaviour, we adapt the local quench construction 
of \cite{Bernard:2012je,Bhaseen:2013ypa} to the case of a non-relativistic fluid with Lifshitz scaling symmetry and 
study the subsequent time evolution for different values of the dynamical critical exponent.  
We begin in Section~\ref{fluids}, where we introduce general properties of such fluids and continue in 
Section~\ref{setupsec} by describing the setup involving a pair of quantum critical heat baths that are brought into 
contact at $t=0$. In Section~\ref{wavesec} we briefly review the theory of shock and rarefaction waves that can appear in 
this context and associated stability conditions. 
In Sections~\ref{rarefactionsec} and \ref{rarefaction_general_z} we apply the general theory to our specific system, 
first for the case of a $z=2$ scale invariant fluid with Galilean boost invariance and then for a general $z\neq 2$ fluid 
without boost symmetry. Finally, we discuss some open questions and possible future directions in Section~\ref{conclusion}.


\section{Perfect fluids with Lifshitz symmetry}
\label{fluids}
For simplicity, below we will focus on the special case of perfect fluids. These are idealised fluids, that are without shear, 
strain or bulk viscosity and do not conduct heat. We begin by introducing the symmetries we will be assuming and the 
definition of the dynamical critical exponent $z$.

\subsection{Symmetries of relativistic and non-relativistic critical fluids}
Symmetries play a central role in any fluid description. The most basic symmetries are time translations, spatial translations 
and spatial rotations, generated by the operators $\mathfrak{g} = \{ \hat H, \hat P_i, \mathcal{\hat J}_{ij}\}$, respectively, 
whose commutators form the so-called Aristotelian algebra.

A relativistic fluid is not only invariant under these symmetries, but also under Lorentz boosts $\hat L_i$ relating 
observers moving with respect to each other with constant velocity,
\begin{equation}
\vec x'=\gamma\,(\vec x-\vec v t), \qquad t'=\gamma\,\left(t-\frac{\vec v \cdot \vec x}{c^2}\right)\,,
\label{lorentz}
\end{equation}
where $\vec v$ describes the relative velocity between the two observers and $\gamma =1/\sqrt{1-v^2/c^2}$. 
At low velocities $v\ll c$, the Lorentz boost reduces to the Galilean boost $\hat G_i$, 
\begin{equation}
\vec x' = \vec x-\vec v\,t, \qquad t'=t\,.
\label{galileo}
\end{equation}
The Aristotelian algebra is extended to the Poincaré algebra or the Galilei algebra, depending on which of these 
boost generators is added to $\mathfrak{g}$. Furthermore, the Galilei algebra allows for a central extension, 
known as the Bargmann algebra \cite{Bargmannold}, by the inclusion of an additional symmetry generator 
$\mathcal{\hat M}$, such that the non-vanishing Galilean boost commutators are given by
\begin{align}
\begin{split}
[\mathcal{\hat J}_{ij}, \hat G_k] &= \hat G_j\,\delta_{ij} - \hat G_i\,\delta_{jk}\,,\\
[\hat H, \hat G_i] &= \hat P_i\,,\\
[\hat P_i, \hat G_j] &= \mathcal{\hat M}\,\delta_{ij}\,.
\label{galileanalgebra}
\end{split}
\end{align}
The charge $\mathcal{\hat M}$ corresponds to the non-relativistic kinetic mass \cite{newtonian} and needs to be 
included when describing a fluid with mass density. In a theory with Galilean boost symmetry, the kinetic mass is a 
measure of the amount of matter in the system and does not vary between inertial frames. It is a conserved quantity 
in an isolated system.

On top of this, in a relativistic \emph{critical} fluid there is an additional symmetry under dilations of the form
\begin{equation}
\vec x' = \Lambda \vec x, \qquad t'=\Lambda t\,,  \quad \textrm{with}\quad\Lambda>0\,.
\label{scaling}
\end{equation}
Invariance under this symmetry implies that physical processes happen in the same way, 
at all distance scales or, alternatively, energy scales. For relativistic fluids, the scale symmetry is 
compatible with Lorentz symmetry and together they place powerful constraints on the allowed dynamics of the fluid. 

A non-relativistic critical fluid can be scale invariant too, but in this case dilations $\mathcal{\hat D}$ take the more 
general form of a Lifshitz symmetry,
\begin{equation}
\vec x' = \Lambda \vec x, \qquad t'=\Lambda^z t\,,
\label{lifshitz}
\end{equation}
where $z\geq 1$ is referred to as the dynamical critical exponent. In the absence of boost symmetries, a closed algebra 
exists for any $z$ consisting of the generators $\mathfrak{g}_z = \{ \hat H, \hat P_i, \mathcal{\hat J}_{ij}, \mathcal{\hat D}\}$.

A key observation, however, is that Lifshitz symmetry with generic $z > 1$ is in general \emph{not} compatible with 
boost symmetry. Indeed, Lorentzian boost symmetry is only compatible with $z=1$, which gives the scaling \eqref{scaling} 
and the no-go result of \cite{10.21468/SciPostPhys.5.1.003} implies that the Galilean boost symmetry 
\eqref{galileo} is only compatible with $z=2$ Lifshitz scaling. In the special case of $z=2$ the Bargmann algebra can 
be be further extended to the Schr\"odinger algebra \cite{Taylor:2015glc} involving the set 
$\{ \hat H, \hat P_i, \mathcal{\hat J}_{ij}, \hat G_i, \mathcal{\hat D}_{(z=2)}\}$.
For this reason, when discussing the out of equilibrium dynamics of non-relativistic fluids, we will 
consider separately the cases $z=2$ and $z\neq 2$, leading to different conclusions about the 
nature of the emergent steady state.


\subsection{Thermodynamics and stress-energy tensor}
\label{fluidtensor}
Based on the considerations above, we will consider a fluid whose description is invariant under 
time and space translations as well as rotations. In addition, we will also assume a global $U(1)$ 
symmetry whose corresponding conserved charge is $N$. This is realized by the basic set of 
generators $\{\hat H, \hat P_i, \mathcal{\hat J}_{ij}, \mathcal{\hat M}\}$. 
Additional symmetries under boosts and rescaling will be considered below.

Global quantities in this fluid include the energy $E$, momentum $\mathbb{\vec P}$, entropy $S$ 
and charge $N$. Locally, we have the energy density $\mathcal{E}=E/V$, momentum density 
$\mathcal{P}_i=\mathbb{P}_i/V$, entropy density $s=S/V$ and charge density $n=N/V$. 
Assuming a configuration where these can be uniformly defined, the fundamental thermodynamic 
relations relating the change of the internal energy to the changes in the rest of the thermodynamic 
state functions are
\begin{equation}
\text{d}E=T\,\text{d}S-P\,\text{d}V+v^i\,\text{d}\mathbb{P}_i+\mu\,\text{d}N\,,
\quad \quad
E=T\,S-P\,V+v^i\,\mathbb{P}_i+\mu\,N  \,,
\label{gibbsduh}
\end{equation}
or, in terms of the associated densities,
\begin{equation}
\text{d}\mathcal{E}=T\,\text{d}s+v^i\,\text{d}\mathcal{P}_i+\mu\,\text{d}n \,,
\quad \quad
\mathcal{E} = T \,s -P +v^i\,\mathcal{P}_i+\mu\, n    \,.
\label{gibbsduh1}
\end{equation}
The thermodynamic forces associated to these parameters are the temperature $T$, 
the pressure $P$, the fluid velocity $\vec v$ and the chemical potential $\mu$.

As argued in \cite{10.21468/SciPostPhys.5.1.003}, assuming a fluid with uniform velocity $\vec v$ 
in the presence of rotational symmetry, the momentum density must be proportional to the only 
directed quantity in the fluid, {\it i.e.} the velocity, 
\begin{equation}
\mathcal{P}_i = \rho\, v_i\,,
\label{prhov}
\end{equation}
and the above thermodynamic relation becomes 
\begin{equation}
\text{d}\mathcal{E}=T\,\text{d}s+v^i\,\text{d}(\rho \, v_i)+\mu\,\text{d}n\,.
\label{gibbsduh2}
\end{equation}
The quantity $\rho$ is referred to as the kinetic mass density. In a theory with Galilean boost 
symmetry it is proportional to the charge density $n$ but in the absence of boost symmetry the 
relation between $n$ and $\rho$ is more complicated.

The dynamical variables enter into the stress-energy tensor of the fluid $T^{\mu}{}_{\nu}$ 
and the current $J^{\mu}$,
whose conservation equations read\footnote{Despite the use of $\mu$, $\nu$ indices, we are not assuming 
	Lorentz symmetry and these indices are not to be raised or lowered using a spacetime metric.} 
\begin{equation}
\partial_\mu T^\mu{}_\nu = 0, \qquad \partial_\mu J^\mu = 0 \,.
\label{conservation}
\end{equation}
Classically the symmetry generators are realised by
\begin{align}
\begin{split}
H & = -\int_V {\rm d}^dx\, T^0{}_0(x)\,,\\
P_i & = \int_V {\rm d}^dx\, T^0{}_i(x)\,,\\
\mathcal{J}_{ij} & = \int_V {\rm d}^dx\, \left(x^iT^0{}_j(x)-x^j T^0{}_i\right)\,,\\
N  & = \int_V {\rm d}^dx\, J^0(x)\,,
\end{split}
\end{align}
which provides direct interpretation for various components of the stress-energy tensor and current. 
In particular, the energy density is $\mathcal{E}=-T^0{}_0$, the momentum density is $\mathcal{P}_i=T^0{}_i$, 
and the charge density is $n=J^0$ in any frame.

For a \emph{perfect fluid} there exists a reference frame, the rest frame, in which there is no momentum density.
The charge current then reduces to just the charge density and the stress-energy tensor involves
only two parameters, the energy density and pressure. Explicitly, in this frame we have
\begin{equation}
{T^\mu}_\nu=\begin{pmatrix}-\mathcal{E}_0 &0\\ 0 & P\,\delta^i_j\end{pmatrix}\,, \qquad J^\mu = (n,0)\,.
\label{restt}
\end{equation}
In any other frame of reference the description will also depend on the velocity $\vec v$ and in the absence 
of boost symmetry the $\vec v$ dependence can be non-trivial. 

If the perfect fluid has Lorentz boost symmetry, the stress-energy tensor and current in the moving frame are 
related to those in the rest frame by a Lorentz boost transformation \eqref{lorentz}.
In Section~\ref{rarefactionsec}, we will be interested in describing a non-relativistic perfect fluid with Galilean boost 
symmetry under \eqref{galileo}. In this case the stress-energy tensor and current in the moving frame are obtained 
from the following transformation rules \cite{DESAXCE2012216},
\begin{equation}
T'^\mu{}_\nu = \frac{\partial x'^\mu}{\partial x^\rho} \frac{\partial x^\sigma}{\partial x'^\nu} \left(T^\rho{}_\sigma 
+ J^\rho\,\Gamma_\sigma\right)\,, \qquad J^\mu = \frac{\partial x'^\mu}{\partial x^\rho} J^\rho \,,
\label{weird}
\end{equation}
where we define $\Gamma_\mu = \left(\frac{1}{2}|v|^2,\vec v \right)$. 
Note that this version of ${T^\mu}_\nu$ does not follow the usual tensor transformation properties, 
because it does not have tensorial status in the context of Galilean relativity. However, it is possible to 
combine ${T^\mu}_\nu$ and $J^\mu$ into an $d \times (d+1)$ dimensional object $\tilde T = (T,J)$ which 
acts as a tensor.\footnote{Due to the last relation in \eqref{galileanalgebra}, which relates the charge operator 
	to a commutator of boosts and spatial translations, the conserved charges should ideally be arranged into a 
	single object, not into two separate ones. For the Poincaré group, we have $[\hat P_i, \hat L_j] = \hat P_0\,\eta_{ij}$, 
	so in the context of special relativity $\tilde T$ automatically decomposes into the tensors ${T^\mu}_\nu$ and $J^\mu$.} 
The conservation equations \eqref{conservation} are merged into one, and spacetime is embedded into a 
higher-dimensional construction of Bargmannian coordinates where a tensorial description arises naturally. 
For an overview of this description in the context of Bargmann theory, see \cite{DESAXCE2012216} and \cite{book}. 

Applying \eqref{weird} to a perfect fluid which is flowing at constant velocity $\vec v$, and described in the rest frame 
by \eqref{restt}, we obtain the following stress-energy tensor and current components \cite{jensen},
\begin{align}
\label{estqcf}
\begin{split}
T^0{}_0 &= -{\mathcal{E}},\\
T^0{}_j &= n\, v_j,\\
T^i{}_0 &=- (\mathcal{E} + P)v^i,\\
T^i{}_j &= P\,\delta^i_j + n\, v^i v_j,\\
J^0 &= n,\\
J^i &= n\, v^i\,;
\end{split}
\end{align}
where $\mathcal{E} = \mathcal{E}_0+\frac{1}{2}n\,v^2$ adds kinetic energy to the energy density. 
From the off-diagonal components we read off the momentum density $\mathcal{P}_i=T^0{}_i=n\,v_i$, 
which fixes the coefficient in \eqref{prhov} to be $\rho=n$.

This last observation can also be obtained from the Ward identity corresponding to Galilean boost symmetry. 
The boost generator can be written as $\hat G_i = t\,\partial_i = G^\mu{}_i \partial_\mu$. 
Due to the non-vanishing Poisson bracket $[\hat P_i, \hat G_j]$ in \eqref{galileanalgebra}, the boost current 
is $b^\mu{}_i = t\,T^\mu{}_i-x_i\,J^\mu$ and the associated Ward identity gives $T^0{}_i=J_i$ \cite{hartongobers}, 
from which $\rho=n$ follows. 
The physical interpretation is that the flow of matter gives rise to momentum density and the inhomogeneous 
term in the transformation of the stress-energy in \eqref{weird} accounts for the addition of momentum density under 
Galilean boosts. 

In Section~\ref{rarefaction_general_z}, where we consider critical fluids with generic $z$, we do not assume any 
boost invariance and the kinetic mass density $\rho$ and the particle number density $n$ are no longer identified 
with each other. Instead, we adopt an ansatz where they appear separately in the stress-energy tensor and the 
current \cite{10.21468/SciPostPhys.5.1.003},
\begin{equation}
T^{\mu\nu}=\begin{pmatrix}-\mathcal{E} & \rho\, v^i\\-(\mathcal{E} + P)\,v^i\;\;&\;\;P\,\delta^{ij} + \rho\, v^i v^j\end{pmatrix}\,, 
\qquad J^\mu =\left(n,n\,v^i\right)\,,
\label{genericstt}
\end{equation}
and then study out of equilibrium evolution. 

The Lifshitz scaling relation \eqref{lifshitz} with $z\neq 1$ implies that space and time coordinates have different scaling 
behavior and this affects how dimensional analysis is carried out.
The energy $E$ is a conserved quantity associated to time translations, so it must scale as the inverse of time, and thus 
the energy density scales as $\mathcal{E}'=\Lambda^{-(d+z)}\mathcal{E}$. 
On the other hand, the individual terms in the thermodynamic relation \eqref{gibbsduh2} must all have the same scaling 
and from there one can infer the scaling behavior of the various thermodynamic variables of the Lifshitz fluid:
\begin{equation}
\begin{array}{llll}
\mathcal{E}' =\Lambda^{-d-z}\mathcal{E}\,,&\quad P' = \Lambda^{-d-z}P\,, 
&\quad T' = \Lambda^{-z}\, T\,, &\quad \mu' = \Lambda^{-z} \mu\,,\\
s' = \Lambda^{-d} \,s\,, &\quad n' = \Lambda^{-d} \, n\,, &\quad \rho' = \Lambda^{-d+z-2} \rho\,, 
&\quad v' =\Lambda^{1-z} \, v\,.
\label{scalingzz}
\end{array}
\end{equation} 
Note that it is only for $z=2$ that the kinetic mass density scales in the same way as the charge density.

The symmetry under Lifshitz scaling \eqref{lifshitz} leads to the Ward Identity, 
$z\,T^{0}_{0} + T^{i}_{i} = 0$, which in turn implies the equation of state
\begin{equation}
d\,P =z\,{\mathcal{E}}-\rho\, v^2\,,
\label{eos}
\end{equation}
where $d$ is the number of spatial dimensions. 
For the particular case of $z=2$, 
the equation of state reduces to $d\,P =2\,{\mathcal{E}}-n\, v^2$ and 
it is easy to see that a Galilean boost of the form $\eqref{weird}$ to 
the rest frame gives the equation of state for a fluid at rest $d\,P=2\,{\mathcal{E}}$. 
However, as mentioned above, scale invariance with generic dynamical critical exponent $z$ is
incompatible with Galilean boost invariance and we will see this explicitly in Section~\ref{rarefaction_general_z}
when we study non-equilibrium steady states of a quantum critical fluid with $z\neq 2$. In this case, the state
variables of a uniformly moving fluid are {\it not\/} equivalent to those of an equilibrium configuration 
viewed in a moving reference frame. 


\section{Local quench between semi-infinite heat baths}
\label{setupsec}
The specific system we consider consists of two semi-infinite heat reservoirs in $d$ spatial dimensions, 
which are brought into contact at time $t=0$ across a flat interface orthogonal to the $x$-coordinate axis.
An equilibrium state of a charged quantum critical fluid is characterized by two energy scales, often taken 
to be the temperature and the chemical potential (due to scale invariance it is only the ratio $T/\mu$ that is 
physically relevant). In the case at hand, we find it convenient to instead use the pressure $P_{L,R}$ and 
charge density $n_{L,R}$ of the two reservoirs to describe the initial state,
\begin{equation}
P(t=0,x) = P_L\, \theta(-x)+P_R\, \theta(x)\,, \qquad
n(t=0,x) = n_L\, \theta(-x)+n_R\, \theta(x)\,,
\label{initialdata}
\end{equation}
and our solution to the resulting fluid dynamical problem will be expressed in terms of the scale invariant 
ratios $P_L/P_R$ and $n_L/n_R$. In what follows, we will consider $P_L/P_R>1$ without loss of generality, 
and arbitrary charge ratio, $0< n_L/n_R < \infty$.

A local quench of this type, with sharp jump functions $\theta(x)$, can serve as a first step towards 
studying out of equilibrium dynamics in a fluid.  The pressure difference between the two reservoirs drives a 
fluid flow between them. One might intuitively expect the sharp initial gradient to be steadily smoothed out with 
the system approaching local equilibrium in the central region, but at the level of leading order hydrodynamics 
this is not the case. Instead, as time evolves, a non-equilibrium steady state (NESS) occupies a growing region 
between the two heat baths, characterised by the presence of a non-zero, constant energy flow, as was 
discussed in \cite{Bernard:2012je,Bhaseen:2013ypa}. 
The properties of the NESS are constrained by the equation of state of the heat baths and the conservation 
of the stress energy tensor and the charge current across the wavefronts, which emanate from the contact region
(see Figure~\ref{fig:waves}). 

An initial value problem in hydrodynamics with piecewise constant initial data, where two fluids 
at equilibrium are joined across a discontinuity, is an example of a so-called Riemann problem \cite{Riemann1860}
in the theory of partial differential equations. A solution, which generically involves shock and rarefaction waves 
propagating outwards from the initial discontinuity, can be found via the techniques described in Section~\ref{wavesec},
allowing the fluid variables that characterise the resulting non-equilibrium steady state to be determined in terms
of the relevant input data. A Riemann problem for a relativistic quantum critical fluid in general dimensions 
was studied in \cite{Bhaseen:2013ypa}. Initially, both outgoing wavefronts were assumed to be shockwaves but it
was later realized \cite{PhysRevD.94.025004,Spillane:2015daa} that above two spacetime dimensions,
a solution with one shockwave and one rarefaction wave is preferred, based on entropy arguments 
and backed by numerical analysis. The existence and universality of the steady state for higher dimensional 
CFTs was studied in \cite{Chang:2013gba}.


\subsection{Formulation of the Riemann problem}
\label{R_prob_sec}
In the present Riemann problem, the heat reservoirs are brought into contact
across a planar surface, that we can take to be orthogonal to the $x$-axis. 
Following \cite{Bernard:2012je,Bhaseen:2013ypa}, we look for a 
solution with wave fronts, traveling in the $x$-direction, that separate space into regions.  
\begin{enumerate}[topsep=8pt,itemsep=4pt,partopsep=4pt, parsep=4pt]
	\item
	A region on the left, with the fluid at rest and stress-energy tensor as in \eqref{restt} with $\mathcal{E}_L$, $P_L$ and $n_L$.
	\item
	Steady state region (or regions) in the middle, with the fluid
	flowing at a {\it constant} flow velocity $\vec v$, and 
	stress-energy tensor as in \eqref{estqcf} with $\mathcal{E}_s$, $P_s$ and $n_s$.
	\item
	A region on the right, with the fluid at rest and stress-energy tensor as in \eqref{restt} with $\mathcal{E}_R$, $P_R$ and $n_R$.
\end{enumerate}
Drawing from the expressions presented in \eqref{estqcf}, in each region
the conservation equations \eqref{conservation} take the following form:
\begin{align}
\begin{split}
\partial_{t}\,\mathcal{E} + \partial_{i} \left( (\mathcal{E}+ P) v^i \right) &= 0, \\
\partial_{t} (\rho v^i) + \partial_{j} (P+\rho v_i v^j) &= 0, \\
\partial_{t}\,n + \partial_j (nv^j) &= 0.
\label{cons}
\end{split}
\end{align}
These equations are supplemented with the equation of state~\eqref{eos} that relates $\mathcal{E}$ and $P$ in a 
way that reflects the scaling symmetry of the fluid system. 

\noindent Thus, the dynamics is governed by a set of hyperbolic conservation laws of the form
\begin{equation}
\partial_{t}\phi + \partial_{i}f = 0,
\label{qfeqs}
\end{equation}
where $\phi$ and $f$ are functions of the same fluid variables and $f(t,x)$ represents the flux 
of the conserved quantity $\phi(t,x)$.
In our non-relativistic quantum critical fluid, the conserved quantities are charge, momentum and energy densities, 
and the resulting conservation equations \eqref{cons} may be written as
\begin{equation}
\partial_{t}\begin{pmatrix}
\mathcal{E}\\ \rho v\\n
\end{pmatrix} = \partial_{x}\begin{pmatrix}
(\mathcal{E}+P)v\\P+\rho v^2\\n v
\end{pmatrix}.
\label{hypercons}
\end{equation}
Let us now discuss briefly the possible wave solutions that will emerge in this system.


\section{Wave analysis}
\label{wavesec}

\begin{figure}[b]
	\includegraphics[width=\linewidth]{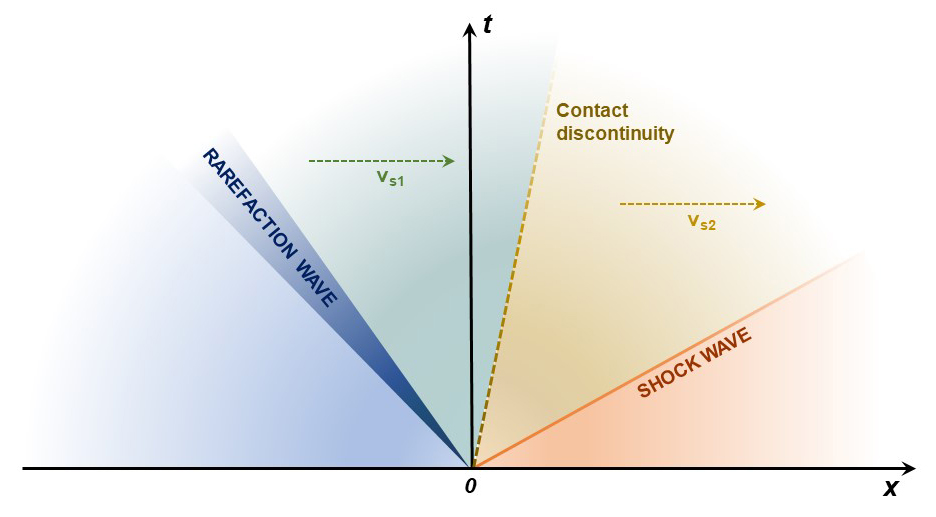}
	\caption{Propagation of shock, contact discontinuity and rarefaction waves for $P_L > P_R$.}
	\label{fig:waves}
\end{figure}

Generically, let us consider a conservation law of the form mentioned above,
\begin{equation}
\partial_t \phi + \partial_x f(\phi) = 0\,,
\label{riemann}
\end{equation}
for a field $\phi(t,x)$, together with a piecewise constant initial condition:
\begin{equation}
\phi(0,x) = \begin{cases}
\phi_L      & \mathrm{if\ } x < 0 \,, \\
\phi_R    & \mathrm{if\ } x > 0 \,.
\end{cases}
\label{ic}
\end{equation}
This problem was first considered by Riemann in the 19th century \cite{Riemann1860}. 
Note that for any given solution of this problem $\phi_{\mathrm{sol}}(t,x)$, 
the rescaled function $\phi_\theta(t,x)=\phi_{\mathrm{sol}}(\theta t,\theta x)$ 
is also a solution for any $\theta>0$.
In fact, the initial condition \eqref{ic} selects, out of all possible solutions of the conservation equations, 
those which are invariant under such a scaling transformation. These solutions are constant along rays 
emanating from the origin $(t=0,x=0)$ due to the scaling, and they can generically be understood in terms 
of waves.


\subsection{Linear problem}
In the problem we will be considering, $\phi$ is a vector whose components are the energy density, 
pressure and fluid velocity, but, for the present discussion, we simply take it to be a generic vector of $k$ 
components. A simple special case is obtained when $\partial_x f(\phi) \propto \partial_x \phi$, that is, for 
the strictly hyperbolic system
\begin{equation}
\partial_t \phi + A\,\partial_x \phi = 0\,,
\end{equation}
where $A$ is a matrix of constant coefficients. In this case, any solution can be written as a superposition 
of traveling waves. A generic initial condition $\phi(0,x)= \hat \phi(x)$ defines a wave profile that is shifted 
to the left and right as it evolves in time, in such a way that the height of the evolved profile at a given point 
is the sum (superposition) of heights at different points of the original profile.

The explicit solution takes the form
\begin{equation}
\vec \phi(t,x)=\sum_{i=1}^k \vec r_i\,\hat \phi_i(x-\lambda_i t),
\label{xtsol}
\end{equation}
where $\lambda_i$ are the eigenvalues of the matrix $A$, that determine the speed of propagation of each 
component of $\vec \phi$, while the coefficients of the superposition, $\vec r_i$, are the components of 
the corresponding eigenvectors of $A$, and they determine the direction of the rays along which the wave 
travels.  By diagonalising the matrix, the problem is decomposed into $k$ scalar Cauchy problems that 
can be solved separately.


\subsection{Non-linear problem}
More generally, the Jacobian in \eqref{riemann} is a function of $\phi$ itself,
\begin{equation}
A(\phi) = 
df(\phi)=
\left.\begin{pmatrix}
\frac{\partial f_1}{\partial \phi_1} &
\cdots &
\frac{\partial f_1}{\partial \phi_k} \\
\vdots &
\cdots &
\vdots \\
\frac{\partial f_k}{\partial \phi_1} &
\cdots &
\frac{\partial f_n}{\partial \phi_k} 
\end{pmatrix}\right. .
\label{jac}
\end{equation}
This adds non-linearity to the problem. The solution can still be written in terms of waves, but the waves 
can interact with each other, producing additional waves. This is because the eigenvectors $r_i$ are 
generalised into functions which depend on $\phi$. The eigenvalues $\lambda_i$ also depend 
on $\phi$, and so the shape of the various components of the solution will vary in time, 
leading to wave dispersion and compression.

In \cite{doi:10.1002/cpa.3160070112}, Lax provided a classification of the waves that can arise in 
non-linear wave problems with initial conditions of the form \eqref{ic}. To do so, he introduced 
a simplifying assumption: that each $\lambda_i(\phi)$, that is, the $i^{\text{th}}$ eigenvalue of the 
Jacobian matrix \eqref{jac}, corresponds to either a \emph{genuinely non-linear} wave, 
such that $\vec \nabla \lambda_i(\phi)\cdot \vec r_i(\phi) \neq 0$ for all $\phi$, 
or to a \emph{linearly degenerate} wave, such that $\vec \nabla \lambda_i(\phi)\cdot \vec r_i(\phi) = 0$ 
for all $\phi$. The quantity $\vec \nabla \lambda_i \cdot \vec r_i$ can be understood as the directional 
derivative of $\lambda_i(\phi)$ in the direction of the vector $\vec r_i$.

As we will see below, this assumption holds in our Riemann problem for Lifshitz fluids and the 
resulting solutions have a simple structure consisting of different kinds of waves or discontinuities, 
which can be classified as follows:

\vspace{12pt}

$\bullet$ The linearly degenerate case $\vec \nabla \lambda_i \cdot \vec r_i = 0$, for which $\lambda_i$ is 
constant along each integral curve of the corresponding field of eigenvectors $r_i$. In this case the profile 
of the solution does not change in time, generating a so-called contact discontinuity.

$\bullet$ The genuinely non-linear case with $\vec \nabla \lambda_i \cdot \vec r_i > 0$ such that the $i^{\text{th}}$
eigenvalue $\lambda_i$ is strictly increasing along the integral curve of the corresponding field of eigenvectors $r_i$. 
This leads to a rarefaction wave, displaying a smooth profile that widens and decays over time.

$\bullet$ The genuinely non-linear case with $\vec \nabla \lambda_i \cdot \vec r_i < 0$. This leads to a shock wave, 
displaying a compression which makes it become steeper over time.

\vspace{12pt}

When the simplifying assumption described above is valid, a set of stability conditions can be 
formulated which guarantee uniqueness and a continuous dependence on the initial data \cite{BIANCHINI2003}. 
The one relevant for our analysis is Lax's shock wave admissibility condition \cite{Lax:1957hec}, which can be 
easily visualised for the Riemann problem, where the initial configuration of $\phi(0,x)$ jumps from a left 
state $\phi_L$ to a right state $\phi_R$ at some value of $x$. 
The information contained in the piecewise initial condition propagates forward at speeds given by 
$\lambda_i(\phi_L)$ on the left and $\lambda_i(\phi_R)$ on the right. In order to prevent new characteristics 
spawning away from the shock interface, which would amount to non-uniqueness for our Cauchy problem, 
one must impose $\lambda_i(\phi_L) \geq \lambda_i(\phi_R)$. 
Furthermore, a shock wave connecting the states $\phi_L$, $\phi_R$ moving at speed 
$\lambda = u_s$, must satisfy
\begin{equation}
\lambda_i(\phi_L) \geq u_s \geq \lambda_i(\phi_R)\,.
\label{lax}
\end{equation}
Lax's admissibility condition applies to shock waves but not to rarefaction waves. For a rarefaction wave, the 
solution's admissibility is determined by requiring $\lambda_i(\phi)$ to increase smoothly along the profile.


\section{Rarefaction and shock waves for a $z=2$ Lifshitz fluid}
\label{rarefactionsec}

As already mentioned in Section~\ref{fluids}, a non-relativistic Lifshitz fluid with scaling exponent $z=2$ is special. 
This is due to a number of reasons. First of all, a Galilean boost invariant field theory describing such a fluid has 
been explicitly  constructed \cite{Hoyos:2013qna,Chemissany:2012du}. In addition, for $z=2$, the Schrödinger 
group (consisting of the Bargmann group, enhanced by the addition of the dilation operator $\mathcal{\hat D}$), 
can have an additional generator, $\hat C$, corresponding to special conformal transformations. 
Finally, as shown in \cite{10.21468/SciPostPhys.5.1.003} and \cite{PhysRevD.97.125006}, it is only for this 
particular value of $z$ that one can have a Galilean boost invariant fluid with Lifshitz scaling symmetry 
with a discrete Hamiltonian and number operator spectrum.

In view of this, we first consider a $z=2$ Lifshitz fluid in $d$ spatial dimensions taken to be invariant under 
Galilean boosts in addition to the scaling symmetry.  In this case, we have the relation $\rho = \,n$ by virtue 
of a Ward identity, so the momentum density \eqref{prhov} is
\begin{equation}
\mathcal{P}_i = n\,v_i\,,
\end{equation}
and the equation of state \eqref{eos} reduces to
\begin{equation}
d\,P = 2\,\mathcal{E}-n\, v^2\,.
\label{z2eos}
\end{equation}
Then, the conservation equations \eqref{hypercons} become
\begin{equation}
\label{hypercons2}
\partial_{t}\begin{pmatrix}[1.5]
\mathcal{E}\\q\\n
\end{pmatrix} 
= \partial_{x}\begin{pmatrix}[1.5]
\frac{(d+2)}{d}\,\frac{q\,\mathcal{E}}{n}-\frac{1}{d}\,\frac{q^3}{n^2}\\
\frac{2}{d}\,\mathcal{E}+\frac{(d-1)}{d}\,\frac{q^2}{n}\\
q
\end{pmatrix},
\end{equation}
where the combination
\begin{equation}
q=n\,v
\label{qdef}
\end{equation}
has been introduced and the right hand side has been 
expressed solely as a function of the conserved variables. This has the form of a Riemann problem \eqref{riemann}
with $\phi = (\mathcal{E},q,n)$. The flux vector $f(\phi)$ can be read off from the right hand side and the
Jacobian matrix is easily evaluated,
\begin{equation}
df(\phi) = 
\begin{pmatrix}[1.8]
\frac{(d+2)}{d}\,\frac{q}{n} & \quad\frac{(d+2)}{d}\,\frac{\mathcal{E}}{n}-\frac{3}{d}\,\frac{q^2}{n^2} 
&\quad -\frac{(d+2)}{d}\,\frac{q\mathcal{E}}{n^2}+\frac{2}{d}\,\frac{q^3}{n^3}\\
\frac{2}{d} & \frac{2(d-1)}{d}\,\frac{q}{n} & -\frac{(d-1)}{d}\,\frac{q^2}{n^2} \\
0 & 1 & 0 \\
\end{pmatrix}\,.
\end{equation}

One of the eigenvalues of the Jacobian, along with the corresponding eigenvector, is 
\begin{gather}
\lambda_1 = \frac{q}{n}, \quad r_1 = \begin{pmatrix}[1.5]\frac{q^2}{2n} \\ q \\ n\end{pmatrix}\,.
\end{gather}
This is linearly degenerate, 
$\vec \nabla\lambda_1(\phi) \cdot \vec r_1(\phi) = 0$,
and corresponds to a contact discontinuity. 

The remaining eigenvalues and eigenvectors are 
\begin{gather}
\label{landa2}
\lambda_2 =  \frac{q}{n}\bigg(1-\frac{\sqrt{d+2}}{d}\sqrt{\frac{2\mathcal{E}n}{q^2}-1}\bigg), 
\qquad r_2 =  \begin{pmatrix}[1.8]
\frac{(d+2)}{d}\,\mathcal{E}n-\frac{1}{d}\,q^2-\frac{\sqrt{d+2}}{d}\,
q^2\,\sqrt{\frac{2\mathcal{E}n}{q^2}-1}\\ 
q\bigg(1-\frac{\sqrt{d+2}}{d}\sqrt{\frac{2\mathcal{E}n}{q^2}-1}\bigg)\\
n\end{pmatrix}\,;
\end{gather}
and
\begin{gather}
\lambda_3 =  \frac{q}{n}\bigg(1+\frac{\sqrt{d+2}}{d}\sqrt{\frac{2\mathcal{E}n}{q^2}-1}\bigg), 
\qquad r_3 =  \begin{pmatrix}[1.8]
\frac{(d+2)}{d}\,\mathcal{E}n-\frac{1}{d}\,q^2+\frac{\sqrt{d+2}}{d}\,
q^2\,\sqrt{\frac{2\mathcal{E}n}{q^2}-1}\\ 
q\bigg(1+\frac{\sqrt{d+2}}{d}\sqrt{\frac{2\mathcal{E}n}{q^2}-1}\bigg)\\
n\end{pmatrix}\,;
\end{gather}
which are genuinely non-linear, {\it i.e.} $\vec \nabla\lambda_i(\phi) \cdot \vec r_i(\phi) \neq 0$.
These two families of solutions correspond to rarefaction and shock waves. Notice that from \eqref{z2eos} 
and \eqref{qdef} it follows that their eigenvalues can be written as $\lambda_2=v-c$ and $\lambda_3=v+c$, 
where $c$ is the local speed of sound in the fluid,
\begin{equation}
c=\sqrt{\frac{(d+2)}{d}\,\frac{P}{n}}. 
\label{cdef}
\end{equation}
It follows that the $\lambda_2$ ($\lambda_3$) eigenvalue corresponds to a left-moving (right-moving) wave.

\begin{center}
	\begin{figure}[t]
		\centering
		\includegraphics[width=.7\linewidth]{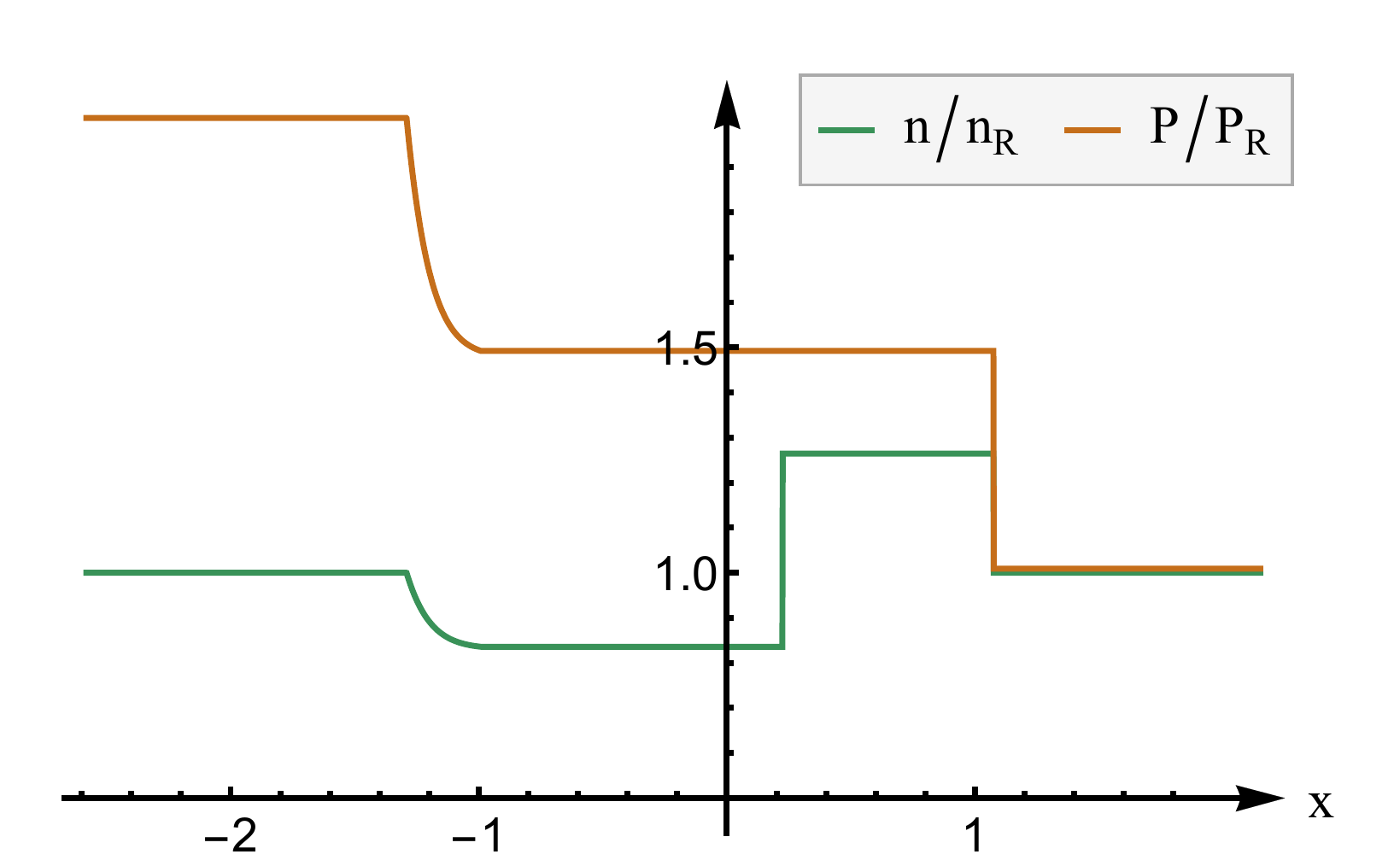}
		\caption{Snapshot of wave profiles for the pressure and charge density at $t=t_0>0$ for $P_L > P_R$ and 
			$n_L=n_R$. The NESS region, bordered by the left-moving rarefaction wave and the right-moving shock wave, contains a contact discontinuity in the charge density.}
		\label{fig:cartoon}
	\end{figure}
\end{center}

Lax's admissibility condition for a shock wave turns out to be satisfied if and only if the pressure in the region 
behind the wave front exceeds the pressure in the region ahead of it. In our problem, where we assume
that $P_L>P_R$, this is the $i=3$ right-moving wave.  The left-moving $i=2$ wave, on the other hand, advances
into a region of higher pressure and is therefore a rarefaction wave, whose profile widens over time.\footnote{Under 
	the reverse assumption, $P_L<P_R$, the only change is that the rarefaction and shock wave profiles are switched
	between the left- and right-moving waves.} Figure~\ref{fig:cartoon} shows a snapshot of the wave profile for a 
particular choice of initial data in \eqref{initialdata}, with a rarefaction wave on the left, a shock wave on the 
right, and a contact discontinuity in between. The shape is similar to the solution of the corresponding  
Riemann problem for a relativistic critical fluid considered in \cite{PhysRevD.94.025004,Spillane:2015daa}.
In particular, as we'll see below, the pressure remains constant across the contact discontinuity in the NESS 
region while the charge density jumps. In the relativistic case, the charge density decouples from the equations 
that determine the pressure but this is not the case here. For a non-relativistic Lifshitz fluid, the pressure still 
remains constant across the contact discontinuity but its value in the NESS region is nevertheless influenced 
by the initial values for the charge density of the two reservoirs (see {\it e.g.} \eqref{Picondition} below).

\subsection{Rarefaction wave profile}
\label{rfwave}
Let us start by analysing the $i=2$ rarefaction wave. For this it is convenient to introduce the concept of 
\emph{Riemann invariants}. A function $R^{(i)}(\phi)$ that is constant along the integral curves of the eigenvector $r_i$,
\begin{equation}
\vec \nabla R^{(i)}(\phi) \cdot \vec r_i = 0,
\label{riemanndef}
\end{equation}
is called an \textit{i}-Riemann invariant. A system with $k$ eigenvalues has $k{-}1$ linearly independent 
\textit{i}-Riemann invariants and they provide a convenient way to construct elementary wave 
solutions that are the building blocks of a full solution to the Riemann problem \cite{Riemann1860}.
In the case at hand, we have two independent Riemann invariants per family of solutions, satisfying
\begin{equation}
\left(\frac{\partial R_A^{(i)}}{\partial \mathcal{E}},\frac{\partial R_A^{(i)}}{\partial q},\frac{\partial R_A^{(i)}}{\partial n}
\right)\cdot \vec r_i = 0\,,\quad \textrm{for} \quad A=1,2 \,.
\label{riemann_inv}
\end{equation}

For the first family, $\lambda_1=\frac{q}{n} = v$ is itself a Riemann invariant, which means that the speed of the 
fluid is the same on both sides of the contact discontinuity, and additionally that the discontinuity itself moves 
at the same constant speed. In fact, this wave is called a contact discontinuity precisely because it 
moves at the fluid flow speed. A second Riemann invariant for the first family is given by the pressure, 
$P= \frac{2\mathcal{E}}{d} -\frac{q^2}{d\,n}$, so this quantity remains constant across the discontinuity as well.

For the two genuinely non-linear families, we find the following pairs of Riemann invariants:
\begin{align}
R_1^{(2)} &= n^{-\gamma}\Big(2\mathcal{E}-\frac{q^2}{n}\Big) \,,
\qquad R_2^{(2)}= \frac{q}{n} + \sqrt{d+2}\sqrt{\frac{2\mathcal{E}}{n}-\frac{q^2}{n^2}}\,, \nonumber \\
R_1^{(3)} &= n^{-\gamma}\Big(2\mathcal{E}-\frac{q^2}{n}\Big) \,,
\qquad R_2^{(3)}= \frac{q}{n} - \sqrt{d+2}\sqrt{\frac{2\mathcal{E}}{n}-\frac{q^2}{n^2}}\,,
\label{rinvariants}
\end{align}
where $\gamma \equiv \frac{d+2}{d}$.
In order to facilitate their interpretation, these expressions can be rewritten using the equation of state, 
\begin{align}
R_1^{(2)} &= n^{-\gamma}P\,,
\qquad R_2^{(2)}= v + d\,c\,, \nonumber\\
R_1^{(3)} &=  n^{-\gamma}P \,,
\qquad R_2^{(3)}= v - d\,c \,,
\label{rinvariants2}
\end{align}
where $c$ was defined in \eqref{cdef} and we have dropped a multiplicative constant from $R_1^{(2)}$ and $R_1^{(3)}$.
We note that $c$ and $\gamma$ are, respectively, the speed of sound and the ratio of specific heats at fixed pressure and 
volume in an ideal gas of $z=2$ Lifshitz particles in $d$ spatial dimensions \cite{10.21468/SciPostPhys.5.1.003}.

The first Riemann invariant is the same for both the second and third families and involves a combination of pressure 
and particle density, $P\,n^{-\gamma}$, which remains constant during an isentropic process in an ideal gas. In other words, 
the conservation of $R_1^{(i)}$ amounts to the conservation of specific entropy, {\it i.e.\/} the entropy per particle, along 
integral curves of $r_i$. To see this, write the first law of thermodynamics in the form 
\begin{equation}
T\,\text{d}s = \text{d}e - \frac{P}{n^2}\text{d}n,
\label{gibbs}
\end{equation}
where $s$ and $e$ are, respectively, the specific entropy and specific internal energy. 
When expressed in terms of the specific internal energy, the equation 
of state \eqref{eos} becomes 
\begin{equation}
d\,P = zne + \frac{z-2}{2}nv^2,
\end{equation}
which reduces to $d\,P = 2ne$ for $z=2$. This implies
\begin{equation}
\text{d}e = \frac{1}{\gamma - 1}\left(\frac1n\text{d}P - \frac{P}{n^2}\text{d}n\right).
\label{equationde}
\end{equation}
Inserting \eqref{equationde} into \eqref{gibbs} and applying the ideal gas law, one obtains
\begin{equation}
\text{d}s = \frac{1}{\gamma - 1}\,\text{d}\log\left(n^{-\gamma}P\right).
\end{equation}
Thus, the first Riemann invariant in \eqref{rinvariants} may be interpreted in terms of entropy and we note that the 
second one has the expected form of a Riemann invariant obtained for a compressible Eulerian fluid \cite{BENARTZI200619}.

For the $i=2$ rarefaction wave, the conservation equations \eqref{hypercons} are solved implicitly 
by the requirement that both Riemann invariants remain constant along the wave profile,
\begin{equation}
R_1^{(2)}(\mathcal{E},q,n) = R_1^{(2)}(\mathcal{E}_L,q_L,n_L), \qquad 
R_2^{(2)}(\mathcal{E},q,n) = R_2^{(2)}(\mathcal{E}_L,q_L,n_L)\,.
\label{r1r2}
\end{equation}
The left reservoir values $\mathcal{E}_L,q_L,n_L$ are realised at the 
leading edge of the rarefaction wave profile and can therefore be taken as a reference.
The above requirement translates into the following two relations:
\begin{equation}
\frac{P_{s1}}{P_L}= \left(\frac{n_{s1}}{n_L}\right)^\frac{d+2}{d}, \qquad v_{s1} = v_L 
+ d\,c_L\left(1-\left(\frac{n_{s1}}{n_L}\right)^\frac{1}{d}\right),
\label{ris2}
\end{equation}
where $v_{s1}$ denotes the fluid flow velocity to the right of the rarefaction wave (see Figure~\ref{fig:waves}) 
and $v_L$ is the fluid flow velocity in the heat bath on the left ($v_L=0$ in a heat bath at rest).
Equivalently, the first relation in \eqref{ris2} can be used to express the flow velocity in terms of pressure 
rather than charge density, 
\begin{equation}
v_{s1} = v_L +d\,c_L\left(1-\left(\frac{P_{s1}}{P_L}\right)^{\frac{1}{d+2}}\right)\,.
\label{vs1eq}
\end{equation}

The phase velocity of the wave is given by the eigenvalue $\lambda_2$, as seen in \eqref{xtsol}, 
which in the present case is given by $\lambda_2 = v - c$ (with $c>0$). 
Taking the wave profile to be parametrised by $n$, the condition for a valid rarefaction wave solution is
\begin{equation}
\lambda_2(\phi(n_L))\leq \lambda_2(\phi(n))\,.
\label{rarefactioncond}
\end{equation}
On the curve we have
\begin{equation}
\lambda_2(n)=v(n) - c(n)=v_L + c_L\left(d-(d+1)\left(\frac{n}{n_L}\right)^\frac{1}{d}\right),
\label{ris3}
\end{equation}
and the rarefaction condition holds provided the charge density is higher in the region ahead of the wave front than 
behind the wave. This is indeed the case when $P_L>P_R$.

Note that since the wave has a smooth profile with spatial dependence $n(x,t)$, the phase velocity of the rarefaction 
wave also acquires a profile, $\lambda_2(x,t)$. On the leading left wavefront, where $n=n_L$, it evaluates to 
$\lambda_2=-c_L$, that is, to the speed of sound in the heat bath on the left. 

Similar considerations apply when $P_L<P_R$, except in this case the rarefaction wave belongs to
the $i=3$ family and moves to the right.


\subsection{Jump conditions and shock wave}
\label{jumpsec}

Riemann invariants are useful when the wave profile is smooth but other methods are needed for dealing with the 
sharp transitions that occur across a shock wave. A solution can be found by imposing so-called Rankine-Hugoniot 
jump conditions \cite{Lax:1957hec, doi:10.1080/00029890.1972.11993023}, which express the conservation laws 
across the wavefront and relate variables in adjacent regions.
For the problem \eqref{qfeqs}, the jump conditions can be stated as
\begin{equation}
u_s[\phi] = [f],
\end{equation}
where $u_s$ is the speed at which the wave front propagates. The symbol $[q]$ indicates a jump in the variable 
$q$ across a front, that is, $[q]=q_R-q_L$.

For our conservation equations \eqref{hypercons}, we get
\begin{align}
\begin{split}
u_s[n] &= [nv],\\
u_s[nv]& = [P+n v^2],\\
u_s[\mathcal{E}] &= [(\mathcal{E}+P)v], \label{eq29}
\end{split}
\end{align}
where $u_s$ is the speed of the wave front in question and $[x]$ denotes the change in the variable $x$ across the
wave front, as described above. Writing $w = v-u_s$ and $\nu = n\,w$, these conditions can 
be expressed as
\begin{align}
[\nu] &= 0,\label{eq30} \\
[P+\nu\,w] &= 0,\label{eq31}\\
[d\,\nu c^2 +\nu\,w^2] &= 0,\label{eq32}
\end{align}
where we have used the equation of state \eqref{z2eos} and the definition $c^2= \gamma\,\frac{P}{n}$.

A trivial and immediate solution is $\nu=[P]=0$, which is the contact discontinuity described by the linearly 
degenerate $i=1$ family of the previous subsection. 
As discussed below \eqref{riemann_inv}, the pressure and fluid speed are the same on both sides of the contact
discontinuity, $P_{s1}=P_{s2}\equiv P_s$ and $v_{s1}=v_{s2}\equiv v_s$,
but in general the energy and particle densities will be discontinuous across the wave front.

A right-moving $i=3$ wave presents a non-trivial solution to the jump conditions.
Assuming that $\nu \neq 0$, we introduce dimensionless variables:
\begin{equation}
\Pi_s \equiv \frac{P_s}{P_R}, \quad y \equiv \frac{n_{s2}}{n_R} = \frac{w_R}{w_s}\,,
\label{auxiliary}
\end{equation}
where the right-most equality follows from the first jump condition \eqref{eq30}. 
The remaining jump conditions \eqref{eq31} and \eqref{eq32} can be re-expressed as
\begin{equation}
\left(\frac{w_R}{c_R}\right)^2 = \frac{y(\Pi_s-1)}{\gamma(y-1)} \qquad\textrm{and}\qquad 
\left(\frac{w_R}{c_R}\right)^2 = \frac{d\,y\,(\Pi_s-y)}{y^2-1} \,,
\label{eq36}
\end{equation}
respectively. Combining these conditions and solving for $y$ or $\Pi_s$ gives
\begin{equation}
y = \frac{(d+1)\Pi_s+1}{d+1+\Pi_s}\qquad \text{or} \qquad \Pi_s=\frac{(d+1)y-1}{d+1-y}\,.
\end{equation}
Substituting $y$ back into \eqref{eq36}, and choosing the branch of the square root
that corresponds to a wave moving to the right, leads to the following expression for the shock speed,
\begin{equation}
u_s = v_R+c_R\sqrt{\frac{1+(d+1)\Pi_s}{d+2}}\,.
\label{ushock}
\end{equation}
Here $v_R$ is the fluid speed in the heat bath on the right ($v_R=0$ for a heat bath at rest).
With this choice of sign, Lax's admissibility conditions \eqref{lax} are satisfied for the shock wave. 
Indeed, with $\lambda_{3, R} = v_R+c_R=c_R$, the requirement is $u_s > c_R$, {\it i.e.} that the speed of the wave 
front exceeds the speed of sound in the medium that the shock wave expands into.  
This, in turn, amounts to the condition $P_s>P_R$. 

Finally, we can use the relation $y=w_R/w_s$ from \eqref{auxiliary} to obtain the fluid speed $v_{s2}$ in the 
region between the shock wave and the contact discontinuity in Figure~\ref{fig:waves},
\begin{equation}
v_{s2}= v_R+c_R\frac{d}{\sqrt{d+2}}
\frac{\left(\Pi_s -1\right)}{\sqrt{(d+1)\Pi_s+1}}\,.
\label{vness}
\end{equation}

\subsection{NESS variables and Galilean boost symmetry}
\label{nessz2}

\begin{figure}[b]
	\begin{subfigure}{0.49\textwidth}
		\includegraphics[width=.96\linewidth]{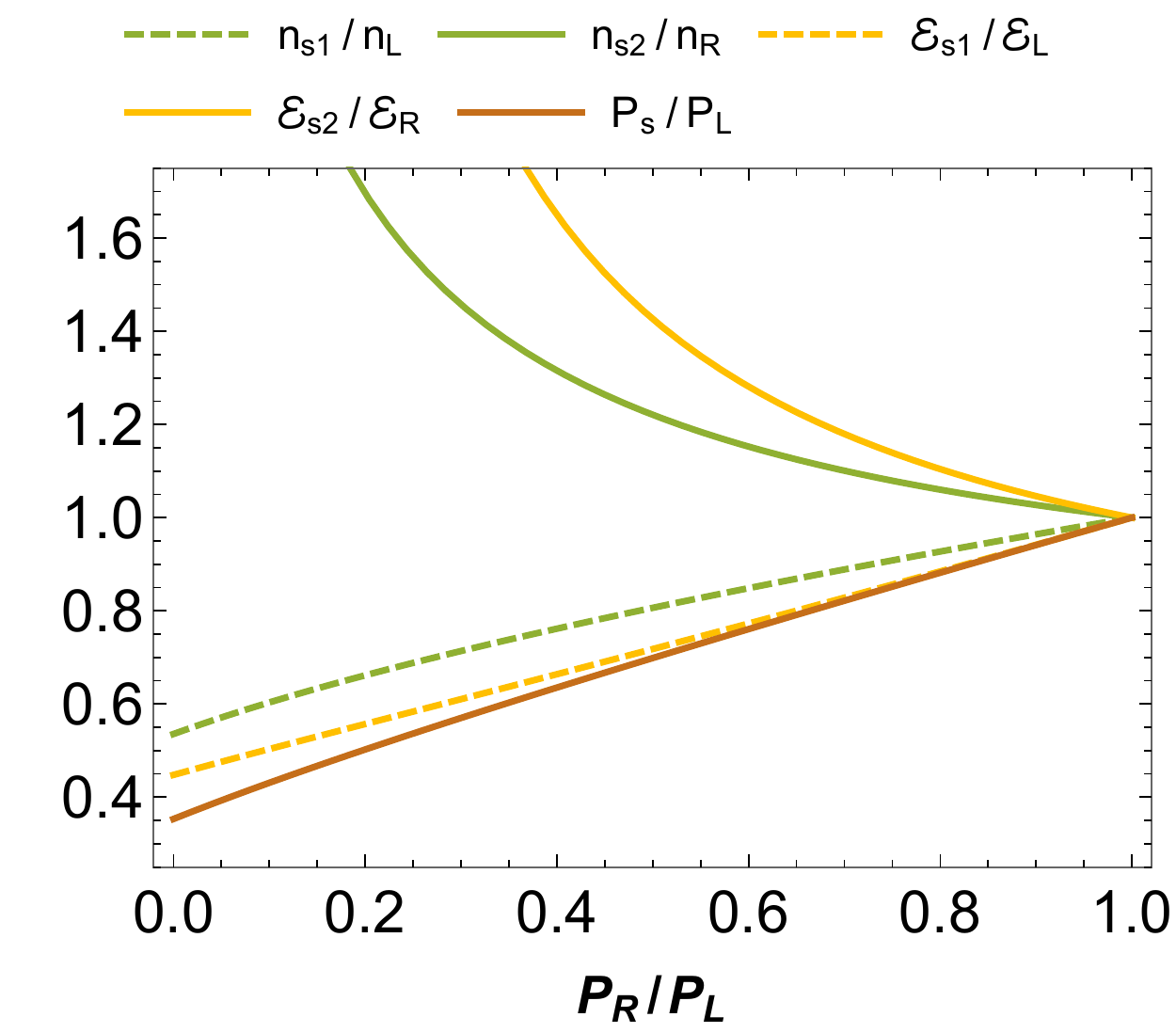}
		\label{fig:Fig 2}
	\end{subfigure}
	\begin{subfigure}{0.49\textwidth}
		\vspace{1.2cm}
		\includegraphics[width=.99\linewidth]{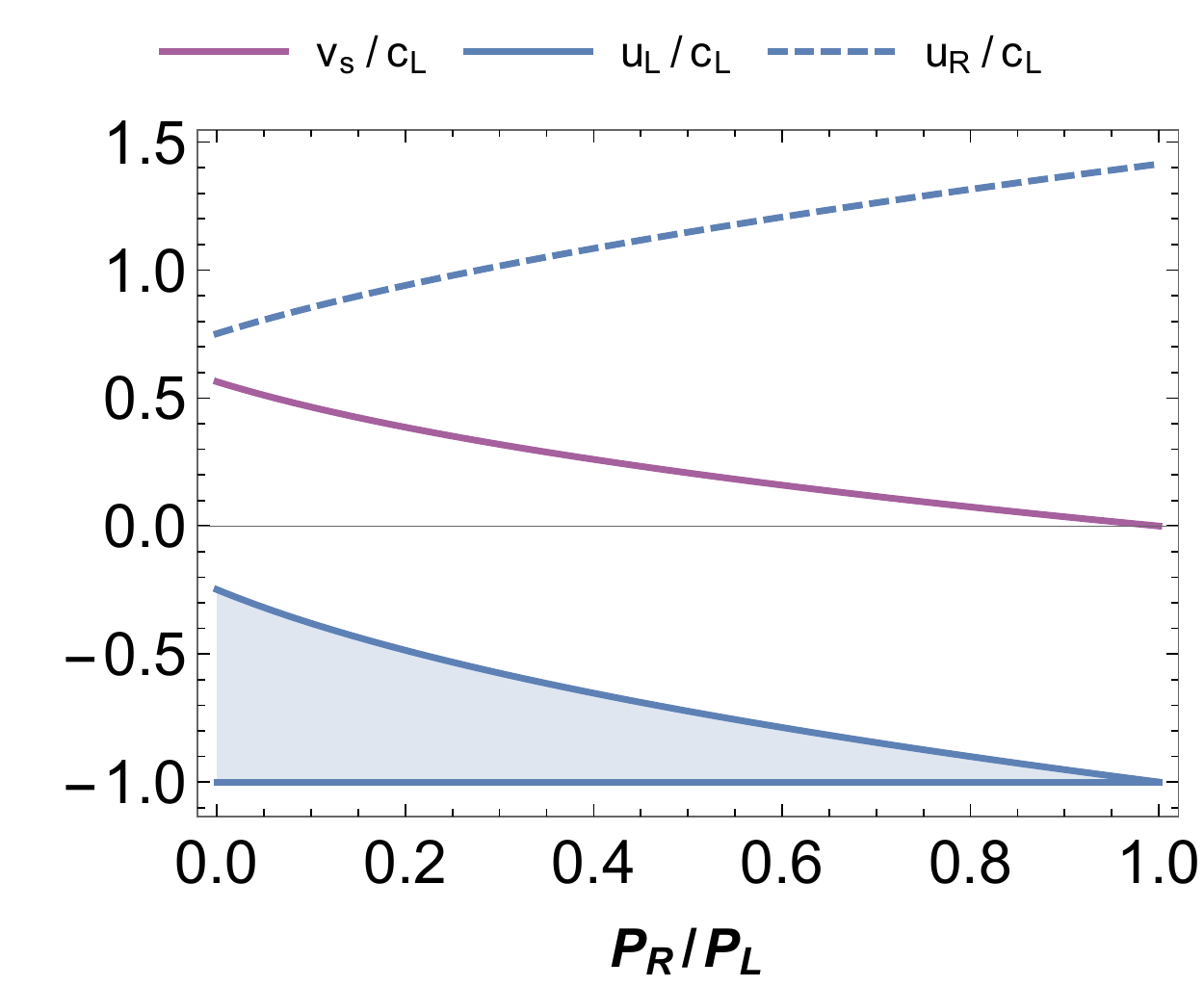}
		\label{fig:Fig 3}
	\end{subfigure}
	\caption{NESS variables for $z=2$, $d=3$ as a function of $P_R/P_L$ for fixed $n_L/n_R=2$. 
		Left panel: Steady state pressure $P_s$, charge densities $n_{s1,s2}$, and energy densities $\mathcal{E}_{s1,s2}$. 
		Right panel: Flow speed $v_s$, shock speed $u_R$, and wave speed $u_L$ across rarefaction profile.}
	\label{fig:Fig 2-3}
\end{figure}

\begin{figure}[t]
	\begin{subfigure}{0.49\textwidth}
		\includegraphics[width=.96\linewidth]{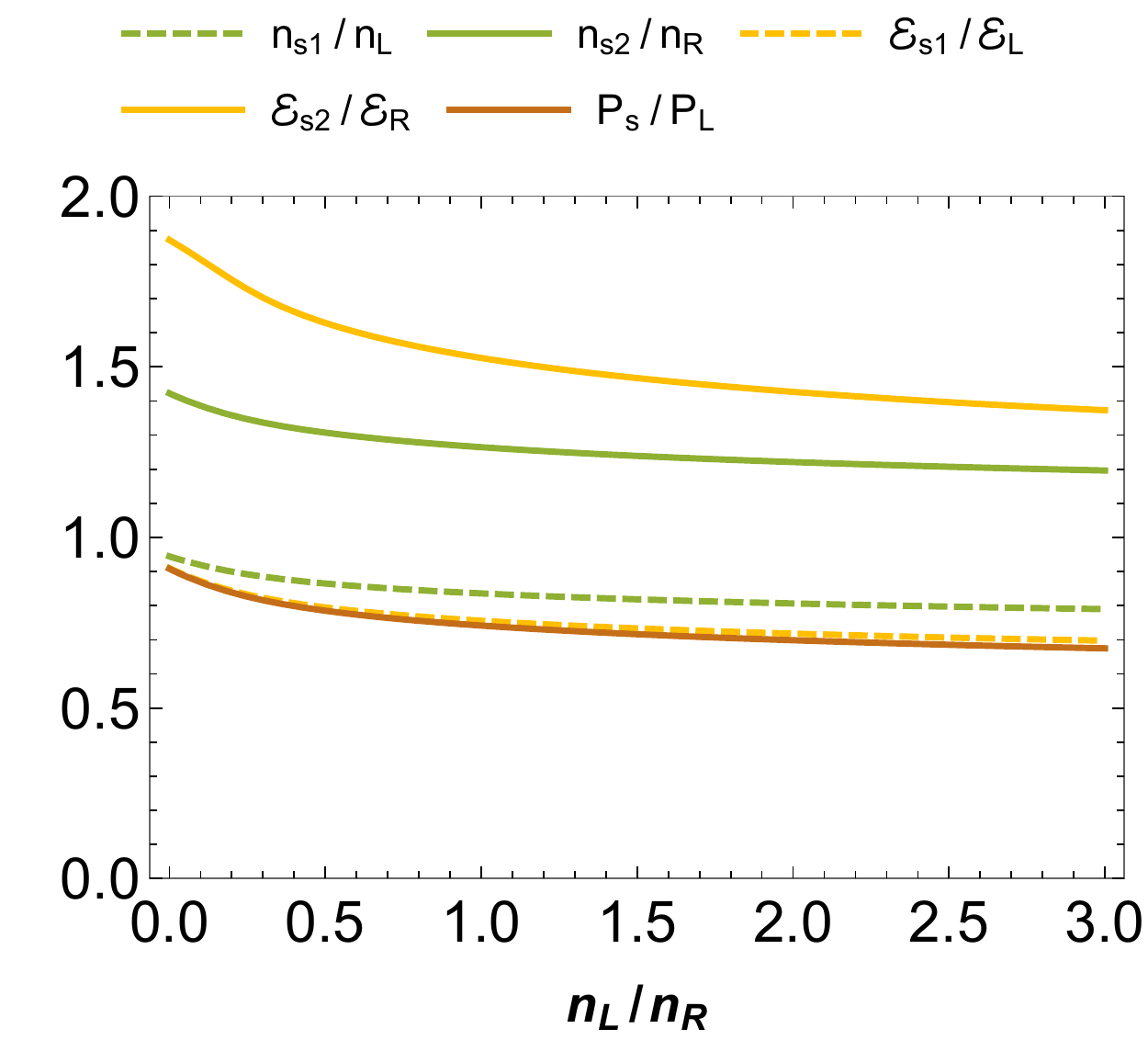}
		\label{fig:Fig 2c}
	\end{subfigure}
	\begin{subfigure}{0.49\textwidth}
		\vspace{0.9cm}
		\includegraphics[width=.99\linewidth]{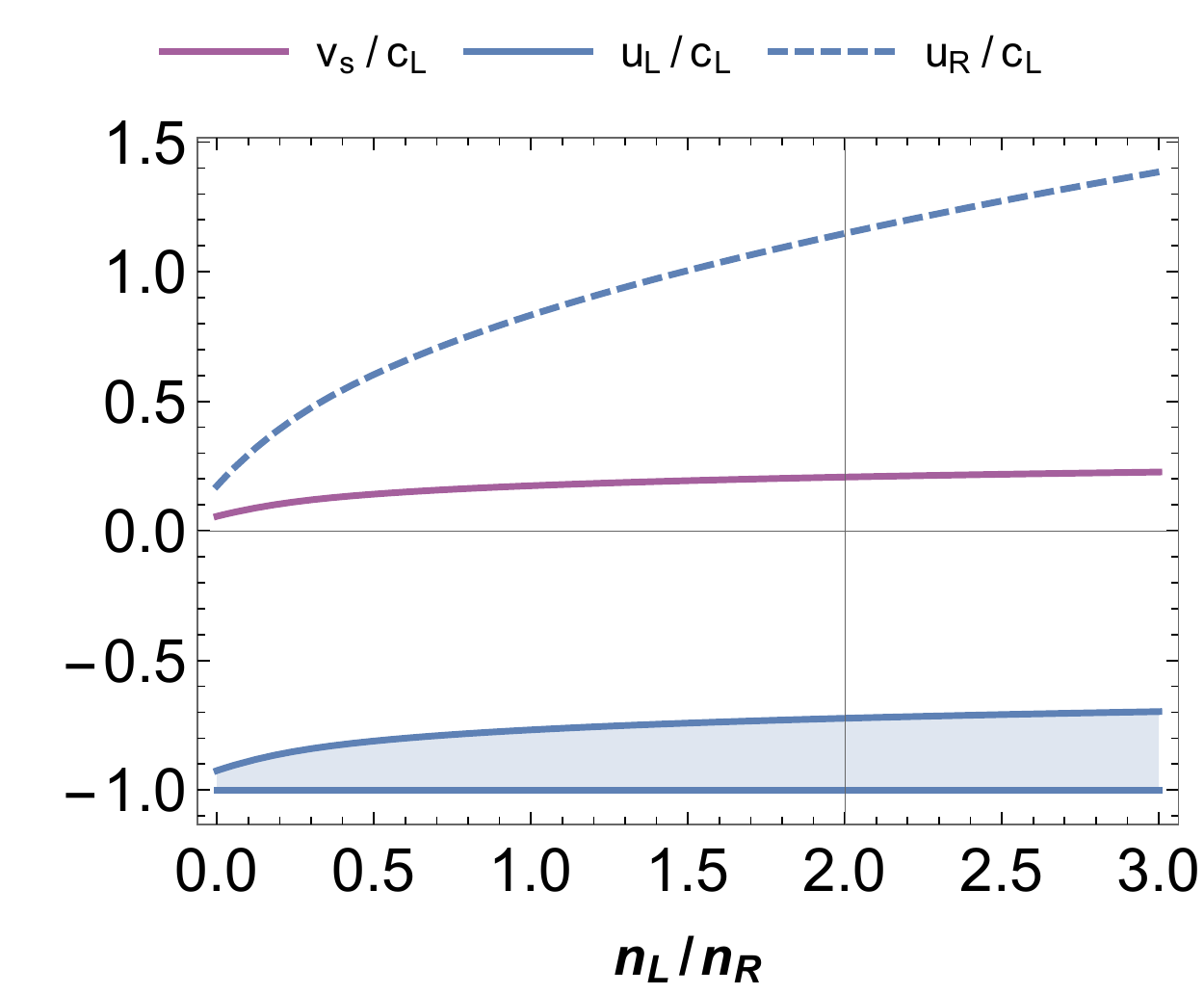}
		\label{fig:Fig 3c}
	\end{subfigure}
	\caption{NESS variables for $z=2$, $d=3$ as a function of $n_L/n_R$ for fixed $P_L/P_R=2$.
		Left panel: Steady state pressure $P_s$, charge densities $n_{s1,s2}$, and energy densities $\mathcal{E}_{s1,s2}$.
		Right panel: Flow speed $v_s$, shock speed $u_R$, and wave speed $u_L$ across rarefaction profile.}
	\label{fig:Fig 2-3c}
\end{figure}

Earlier we observed that pressure and fluid flow speed are the same on both sides of a contact discontinuity 
and the discontinuity itself propagates at the same speed.
Demanding equality of the expressions for $v_{s1}$ in \eqref{vs1eq} and $v_{s2}$ in \eqref{vness} gives us 
the following scale invariant condition on the pressure in the NESS region between the rarefaction and shock waves,
\begin{equation}
1-\left(\frac{\Pi_s}{\Pi_L}\right)^{\frac{1}{d+2}}=
\frac{1}{\sqrt{d+2}}\, \sqrt{\frac{\eta\>}{\Pi_L}}\,\frac{\Pi_s-1}{\sqrt{(d+1)\Pi_s+1}}\,.
\label{Picondition}
\end{equation}
The initial data of the two reservoirs enters through the ratios 
$\Pi_L=P_L/P_R$ and $\eta=n_L/n_R$. The above condition is non-linear but can be solved numerically
and one finds a unique value of $\Pi_s$ for given $\Pi_L$ and $\eta$. The full solution to the Riemann 
problem can then be mapped out by evaluating the following expressions for the remaining NESS 
variables in terms of the pressure,
\begin{align}
\frac{n_{s1}}{n_L} &=\Big(\frac{\Pi_s}{\Pi_L}\Big)^{\frac{d}{d+2}}\,, \label{ns1eq} \\
\frac{n_{s2}}{n_R} &=\frac{(d+1)\Pi_s+1}{d+1+\Pi_s} \,, \label{ns2eq} \\
\frac{\mathcal{E}_{s1}}{\mathcal{E}_L} &=  
\frac{\Pi_s}{\Pi_L} \Big(1+(d+2) 
\Big(\Big(\frac{\Pi_L}{\Pi_s}\Big)^\frac{1}{d+2}-1\Big)^2\,\Big) , \label{Es1eq} \\
\frac{\mathcal{E}_{s2}}{\mathcal{E}_R} &=\Pi_s+\frac{(\Pi_s-1)^2}{d+1+\Pi_s}\, , \label{Es2eq} 
\end{align}
and evaluating \eqref{ushock} for the speed of the right-moving shock wave. The speed of the fluid flow in
the NESS region can be obtained by evaluating either \eqref{vs1eq} or \eqref{vness}. 
Solutions for $d=3$ spatial dimensions are presented in Figure~\ref{fig:Fig 2-3} as a function of $P_R/P_L$
for fixed $n_L/n_R$ and in Figure~\ref{fig:Fig 2-3c} as a function of $n_L/n_R$ for fixed $P_R/P_L$.

In the solution of the corresponding Riemann problem for a relativistic quantum critical 
fluid \cite{PhysRevD.94.025004,Spillane:2015daa} the NESS was described by a Lorentz boosted 
thermal state with a contact discontinuity in the charge density in the fluid rest frame. 
The behaviour of a $z=2$ non-relativistic critical fluid is analogous, although in this case the
boost symmetry is Galilean rather than Lorentzian. The fluid variables in the NESS region of the $z=2$ flow have a 
stress-energy tensor and current of the form \eqref{estqcf}. The pressure and fluid speed are the same on both sides 
of the contact discontinuity but the energy density and the charge density take different values on the two sides. 
Nevertheless, if we perform a Galilean boost with velocity $-v_s$ to the NESS rest frame following the rule \eqref{weird}, 
we obtain a stress-energy tensor of the form \eqref{restt} with $P=P_s$ and a {\it uniform\/} energy density, 
$\mathcal{E}_0=\mathcal{E}_{s1}-\frac12 n_{s1}v_s^2=\mathcal{E}_{s2}-\frac12 n_{s2}v_s^2$. 
Furthermore, the fluid variables in the rest-frame satisfy the equation of state of $z=2$ fluid at rest, $\mathcal{E}_0=\frac{d}{2}P$.
Since $n$ does not transform under a Galilean boost, there is still a contact discontinuity in the charge density.  
Indeed, in the NESS rest frame the two fluids are at rest in hydrostatic equilibrium but the charge density is discontinuous 
across the contact surface.
The charge density remains unchanged with time as there is no fluid flow across the boundary and therefore no charge transport. 
This kind of a sharp charge discontinuity is allowed when we restrict ourselves to leading-order hydrodynamics but is presumably 
smoothed out by higher-order corrections, which we do not consider here. We note that analogous behaviour was seen in the 
NESS rest frame of a relativistic fluid in \cite{PhysRevD.94.025004,Spillane:2015daa}.


\section{Rarefaction and shock waves at general $z$}
\label{rarefaction_general_z}

In this section we turn our attention to a perfect Lifshitz fluid with a general dynamical critical exponent $z>1$.
This is motivated by the existence of quantum critical condensed matter systems with a general dynamical critical 
exponent $z \neq 2$, such as the heavy fermion metals discussed in \cite{articlee} and \cite{PhysRevB.90.045105}.
For generic values of $z$ such a system is without boost symmetry and it is interesting to see how this
affects the solution to the fluid Riemann problem that we have been considering. 
The first thing to note is that the kinetic mass density $\rho$ can no longer be proportional to the charge density $n$ 
when $z\neq 2$. If we assume that $\rho$ can still be expressed as a function of $n$ alone, then the 
scaling relations \eqref{scalingzz} imply a relationship of the form 
\begin{equation}
\rho = m\, n^\alpha \,,
\label{newrho}
\end{equation}
with $\alpha=\frac{d+2-z}{d}$ and $m$ a constant of proportionality. In principle, one could allow for more general 
behaviour, for instance by letting $\rho$ depend explicitly on the velocity $v$ as well as on the charge density, but we will not pursue this here. A scaling ansatz of the form \eqref{newrho} provides an example of 
a Lifshitz fluid without boost symmetry and this is sufficient for our present purposes.
In what follows, we will take $m=1$ for simplicity.

With the above ansatz the thermodynamic relation \eqref{gibbsduh2} takes the form
\begin{equation}
\text{d}\mathcal{E}=T\,\text{d}s+\frac{1}{2}\,n^\alpha \text{d}v^2+\left(\mu+\alpha \,n^{\alpha-1} v^2\right)\,\text{d}n\,.
\label{gibbsduh3}
\end{equation}
The $\text{d}v^2$ terms can be absorbed by defining an internal energy and a shifted chemical potential,
\begin{equation}
\hat{\mathcal{E}}=\mathcal{E}-\frac{m}{2}n^\alpha v^2\,,
\quad \hat{\mu}=\mu+\alpha\,n^{\alpha-1} v^2\,,
\label{redeff}
\end{equation}
and then the familiar form of the first law of thermodynamics is recovered, 
\begin{equation}
\text{d}\hat{\mathcal{E}}=T\,\text{d}s+\hat{\mu}\,\text{d}n\,.
\label{gibbsduh4}
\end{equation}
The equation of state \eqref{eos} becomes
\begin{equation}
d\,P = z\mathcal{E}-n^\alpha v^2\,,
\label{neweos}
\end{equation}
and the conservation equations \eqref{hypercons} can be expressed
\begin{equation}
\label{cons_z}
\partial_{t}\begin{pmatrix}[1.5]
\mathcal{E}\\q\\n
\end{pmatrix} = \partial_{x}\begin{pmatrix}[1.3]
\frac{d+z}{d}\, q\mathcal{E}n^{-\alpha}-\frac{1}{d}\, q^3 n^{-2\alpha}\\
\frac{z}{d}\,\mathcal{E}+\frac{d-1}{d}\, q^2 n^{-\alpha}\\
q \,n^{1-\alpha}
\end{pmatrix},
\end{equation}
with $q=n^\alpha\,v$. The analysis of the Riemann problem proceeds along the same lines as before.
The equations are more involved when $z\neq 2$, and we have to rely on numerical evaluation to a greater
extent, but the NESS variables can still be solved for.

The Jacobian matrix, $df(\phi)$ for general $z$ is 
\begin{equation}
df(\phi) = 
\begin{pmatrix}[1.8]
\frac{d+z}{d}\,q n^{-\alpha} \ \ \ & \>\frac{d+z}{d}\,\mathcal{E}n^{-\alpha}-\frac{3}{d}\,q^2 n^{-2\alpha} 
\ \ \ & \> -\frac{\alpha(d+z)}{d}\,q\mathcal{E}n^{-1-\alpha}+\frac{2\alpha}{d}\,q^3n^{-1-2\alpha} \\
\frac{z}{d} & \frac{2(d-1)}{d}\,q n^{-\alpha} & -\frac{\alpha(d-1)}{d}\,q^2 n^{-1-\alpha} \\
0 & n^{1-\alpha} & (1{-}\alpha)\,q\, n^{-\alpha} \\
\end{pmatrix}\,,
\end{equation}
and its eigenvalues and eigenvectors can readily be evaluated. 
They correspond to a linearly degenerate wave,
\begin{gather}
\lambda_1 = q\,n^{-\alpha}, \quad r_1 
= \begin{pmatrix}[1.5]\frac{\alpha}{z}\, q^2n^{-\alpha} \\  \alpha \,q \\ n\end{pmatrix}\,,
\end{gather}
which is a contact discontinuity, together with two genuinely non-linear waves,
\begin{gather}
\lambda_2 = q\,n^{-\alpha}\left(1+\frac1d(z-2-K)\right) , \qquad
r_2 = \begin{pmatrix}[1.8]
\frac{d+z}{d}\mathcal{E}-\frac{1}{d}\,q^2n^{-\alpha}\left(1+K\right) \\
q(1-\frac{1}{d}\,K) \\ n\end{pmatrix}\,;
\end{gather}
and
\begin{gather}
\lambda_3 =  q\,n^{-\alpha}\left(1+\frac1d(z-2+K)\right) , \qquad 
r_3 =  \begin{pmatrix}[1.8]
\frac{d+z}{d}\mathcal{E}-\frac{1}{d}\,q^2n^{-\alpha}\left(1-K\right) \\
q(1+\frac{1}{d}\,K) \\ n\end{pmatrix}\,;
\label{lambda3_z}
\end{gather}
where we've introduced the shorthand notation, 
\begin{equation}
K\equiv\sqrt{(d+z)\left(\frac{z\mathcal{E}n^\alpha}{q^2}-1\right)-(z-2)} \,.
\label{Kdef}
\end{equation}
It is easily checked that the corresponding expressions in Section~\ref{rarefactionsec} are recovered when we insert $z=2$ in 
\eqref{cons_z} - \eqref{Kdef}. Furthermore, by using the equation of state \eqref{neweos} one obtains
\begin{equation}
K= d\sqrt{\frac{c^2}{v^2}-\frac{(z-2)}{d^2}}, \qquad \textrm{with}\qquad c=\sqrt{\frac{(d+z)}{d}\frac{P}{n^\alpha}}\,.
\label{Kvalue}
\end{equation}
The eigenvalues corresponding to genuinely non-linear waves can then be written,
\begin{align}
\lambda_2&=v\left(1+\frac{z-2}{d}\right)-\sqrt{c^2-\frac{(z-2)}{d^2}v^2} \,, \nonumber \\
\lambda_3&=v\left(1+\frac{z-2}{d}\right)+\sqrt{c^2-\frac{(z-2)}{d^2}v^2} \,,
\end{align}
As before, we find that $\lambda_2$ ($\lambda_3$) corresponds to a left-moving (right-moving) wave, 
and that the leading wavefront of a rarefaction wave will advance at the speed of sound 
in a heat bath at rest.

\subsection{Rarefaction wave profile}
\label{rfprofile}

Now consider initial data of form \eqref{initialdata} for a Lifshitz fluid with general $z$ 
and assume that $P_L>P_R$. In parallel with the $z=2$ case considered in Section~\ref{rarefactionsec}, 
this results in a left-moving rarefaction wave, a right-moving shock wave, and a central
NESS region with constant flow velocity and a contact discontinuity moving with the fluid. 
The key difference compared to the $z=2$ case is that now there is no boost symmetry
and the steady state flow in the central region will no longer be a boosted thermal state. 

We use Riemann invariants to analyse the $i=1$ contact discontinuity and the $i=2$ rarefaction wave. 
The Riemann invariants for the first family of wave solutions are again given by the pressure $P$ 
and the velocity $v$, which coincides with the eigenvalue $\lambda_1 = \frac{q}{n^\alpha}$. 
Therefore, the contact discontinuity will still propagate at the same speed as the velocity of its 
surrounding fluid regions on the left and right.

For the genuinely non-linear families, we find generalisations of the pairs of Riemann invariants, which took the 
form \eqref{rinvariants2} for $z=2$, but are now given by
\begin{align}
\begin{split}
\label{risz}
R^{(2)}_1 =n^{-\gamma}\left(P-\frac{(z-2)}{2d}v^2n^\alpha\right)\,, \qquad 
R^{(2)}_2 =n^{-\xi}\,v\,(1+K)\left(\frac{K-\beta}{K+\beta}\right)^\frac{\beta}{2} \,, \\
R^{(3)}_1 =n^{-\gamma}\left(P-\frac{(z-2)}{2d}v^2n^\alpha\right)\,, \qquad 
R^{(3)}_2 =n^{-\xi}\,v\,(1-K)\left(\frac{K+\beta}{K-\beta}\right)^\frac{\beta}{2} \,,
\end{split}
\end{align}
where $\gamma=\frac{d+z}{d}$, $\xi=\frac{(z-2)(d+z)}{2d}$,  $\beta=\sqrt{\frac{(z-2)(d+z-2)}{2}}$, 
and $K$ may be read off from \eqref{Kvalue}.

\begin{figure}[t]
	\centering
	\includegraphics[width=.52\linewidth]{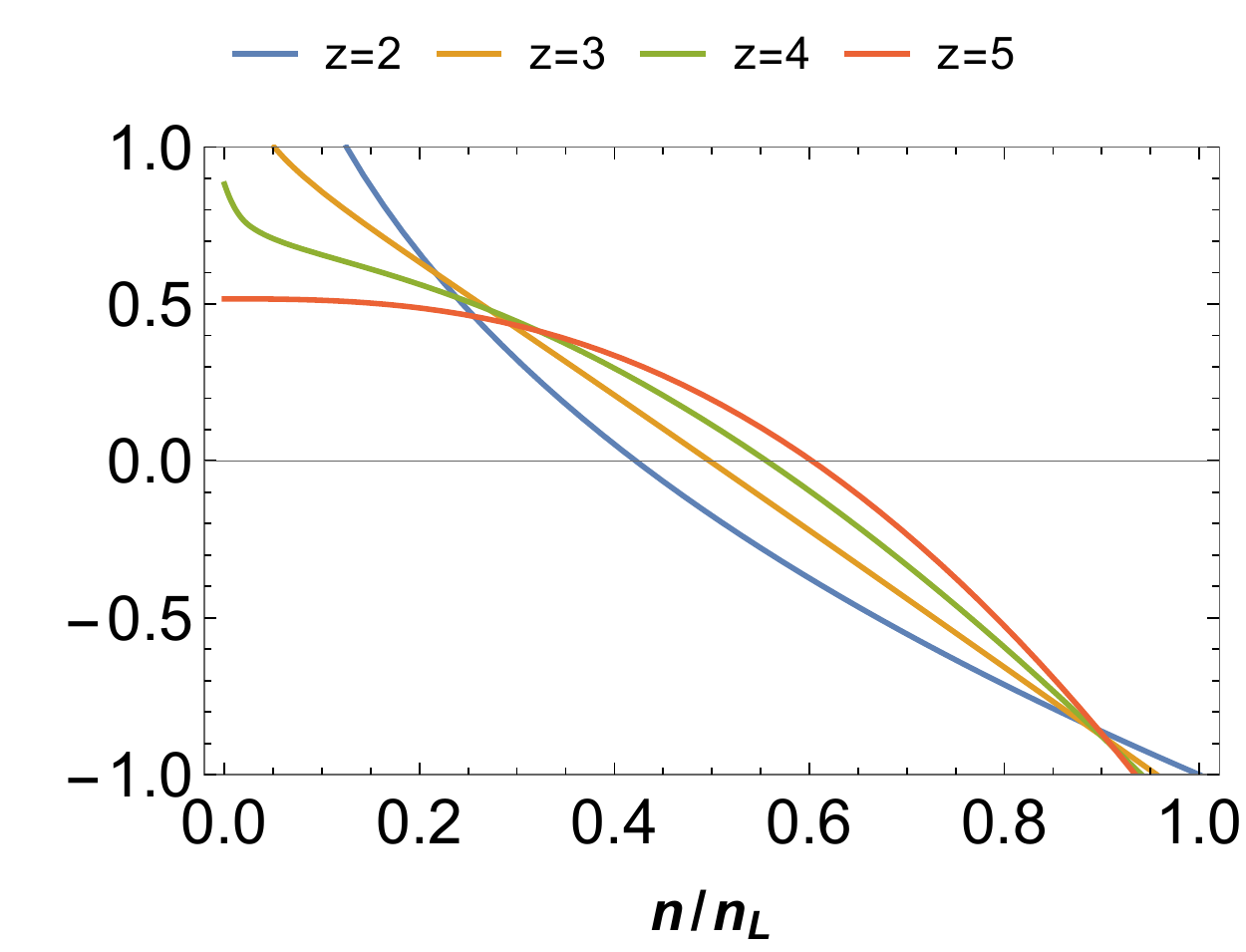}
	\caption{Variation of the characteristic speed $\lambda_2$ across a rarefaction wave profile
		parametrised by $n$ for $d=3$ and different values of $z$.}
	\label{fig:FigXX}
\end{figure}

As before, we require that both Riemann invariants are constant along the characteristic curves of
the left-moving rarefaction wave. From $R^{(2)}_1(P, v, n) = R^{(2)}_1(P_L, v_L, n_L)$ we obtain
\begin{equation}
\frac{P}{P_L}=\frac{\xi}{d}\frac{v^2}{c_L^2}\left(\frac{n}{n_L}\right)^\alpha+\left(\frac{n}{n_L}\right)^\gamma \,,
\label{solprf}
\end{equation}
while $R^{(2)}_2(P, v, n) = R^{(2)}_2(P_L, v_L, n_L)$ gives
\begin{equation}
\left(\frac{n}{n_L}\right)^\xi=\frac{v}{d\, c_L}\left(1+K\right) \left(\frac{K-\beta}{K+\beta}\right)^\frac{\beta}{2} \,,
\label{solnrf}
\end{equation}
with $K$ expressed as a function of $P$, $v$, and $n$ through the relations in \eqref{Kvalue}. 
These conditions are non-linear and do not allow for analytic solution for generic values of $d$ and $z$.
In order to facilitate their numerical solution, we find it convenient to first eliminate the pressure between them
by inserting \eqref{solprf} into \eqref{solnrf}.  This results, after some algebraic manipulations, in 
the following equation, relating the scale invariant variables $n/n_L$ and $v/c_L$,
\begin{equation}
\left(\frac{n}{n_L}\right)^\xi=\left(\frac{v}{d\, c_L}+\widetilde{K}\right)
\left(\frac{\widetilde{K}-\frac{\beta}{d}\frac{v}{c_L}}{\widetilde{K}+\frac{\beta}{d}\frac{v}{c_L}}\right)^\frac{\beta}{2}\,,
\label{solvrf}
\end{equation}
with $\widetilde{K}=\sqrt{\frac{\beta^2}{d^2}\frac{v^2}{c_L^2}+\left(\frac{n}{n_L}\right)^{\gamma-\alpha}}$.
A numerical solution for $v/c_L$ in terms of $n/n_L$ can then be inserted into \eqref{solprf} to determine
$P/P_L$.  In order to check the validity of the rarefaction wave solution so obtained, we have evaluated the 
characteristic speed $\lambda_2$ along the integral curve for specific initial data. Numerical results for several 
different values of $z$ are shown in Figure~\ref{fig:FigXX} and in each case the rarefaction condition \eqref{rarefactioncond}
is indeed satisfied.


\subsection{Shock wave}

A shock wave solution for a Lifshitz fluid at general $z$ satisfies the following Rankine-Hugoniot jump conditions, 
\begin{align}
\begin{split}
u_s[n] &= [nv]\,,\\
u_s[n^\alpha v] &= [P + n^\alpha v^2]\,,\\
u_s[\mathcal{E}] &= [(\mathcal{E}+P)v]\,.
\end{split}
\end{align}
Writing $w = v - u_s$ and $\nu = n^\alpha\, w$, these conditions can be expressed as
\begin{align}
\begin{split}
[n\,w] &= 0\,,\\
[P+\nu(w+u_s)] &= 0\,,\\
[(d+z)Pw+\nu(w+u_s)(w+u_s-zu_s)] &= 0\,,
\end{split}
\end{align}
where we have used the equation of state \eqref{neweos}. 
The contact discontinuity corresponds to the trivial solution $w=[P]=0$. 

To find a right-moving shock wave solution corresponding to the $i=3$ eigenvalue family, we again introduce 
dimensionless variables,
\begin{equation}
\Pi_s \equiv \frac{P_s}{P_R}, \quad y \equiv \frac{n_{s2}}{n_R} = \frac{w_R}{w_s},
\label{auxiliaryz}
\end{equation}
where the right-most equality follows from the jump condition $[n\,w] = 0$. For a shock wave propagating into 
a fluid at rest, the other two jump conditions can be re-expressed as 
\begin{equation}
\frac{u_s^2}{c_R^2} = \frac{y^{2-\alpha}(\Pi_s-1)}{\gamma\,(y-1)} \qquad \textrm{and}\qquad
\frac{u_s^2}{c_R^2} = \frac{d\, y^{2-\alpha}(\Pi_s-y)}{(y-1)(1+(z-1)y)} \,.
\label{jump_rewrite}
\end{equation}
The two equations can now be combined and solved either for $y$ or $\Pi_s$,
\begin{equation}
y = \frac{(d+z-1)\Pi_s+1}{d+1+(z-1)\Pi_s} \,, \qquad  \Pi_s=\frac{(d+1)y-1}{d+z-1-(z-1)y}\,.
\label{yandPsol}
\end{equation}
By substituting $\Pi_s$ into \eqref{jump_rewrite}, the speed of the shock wave can be written in terms of 
the dimensionless variable $y$ as,
\begin{equation}
u_s =c_R \sqrt{\frac{d\, y^{2-\alpha}}{d+z-1-(z-1)y}}\,.
\label{shockspeed_z}
\end{equation}
The shock wave admissibility conditions are satisfied when the shock front moves faster than the speed of sound
in the medium the wave is expanding into, {\it i.e.} when $u_s>c_R$. It is easily checked that this holds for all values 
of $y$ that correspond to $P_s>P_R$.

The fluid velocity in the region between the shock wave and the contact discontinuity can also be expressed in 
terms of $y$ via the relation,
\begin{equation}
v_{s2} = \frac{(y-1)}{y}\,u_s \,.
\label{nessflowspeed}
\end{equation}


\subsection{NESS variables}

We now have everything in place to construct the full solution to our Riemann problem for a Lifshitz fluid with 
general $z$ in $d$ spatial dimensions, with initial data given by $P_L$, $P_R$, $n_L$, and $n_R$ (with $P_L\geq P_R$). 
Once again, there will be a growing NESS region between a left-moving rarefaction wave and a
the right-moving shock wave, with a contact discontinuity in between, as depicted in Fig.~\ref{fig:waves}. 
The solution can be constructed in a number of ways but the key observation is that pressure and fluid flow
speed remain constant across the entire NESS region, while the charge density is piecewise constant and 
makes a jump at the contact discontinuity. We will proceed by first solving for the pressure and flow speed 
in terms of the charge density on either side of the contact discontinuity. We then require that the results are the
same on both sides and this, in turn, fixes the charge densities in terms of the initial data.

On the one hand, the NESS pressure and flow speed are expressed in terms of the dimensionless variable 
$y= n_{s2}/n_R$ in \eqref{yandPsol} and \eqref{nessflowspeed}, respectively. These relations follow directly 
from the Rankine-Hugoniot jump conditions across the shock wave front.

On the other hand, we can obtain the same quantities in terms of another dimensionless variable 
$x=n_{s1}/n_L$ by considering the trailing end of the rarefaction wave profile, where $n=n_{s1}$. 
In this case, \eqref{solvrf} reduces to 
\begin{equation}
x^\xi=\left(\frac1d \frac{v_s}{c_L}+\widetilde{K}_s\right)
\left(\frac{\widetilde{K}_s-\frac{\beta}{d}\frac{v_s}{c_L}}{\widetilde{K}_s+\frac{\beta}{d}\frac{v_s}{c_L}}\right)^\frac{\beta}{2}\,,
\label{solvx}
\end{equation}
with $\widetilde{K}_s=\sqrt{x^{\gamma-\alpha}+\frac{\beta^2}{d^2}\frac{v_s^2}{c_L^2}}$. This  
can be solved numerically for $v_s$ as a function of $x$ and the result is then inserted into \eqref{solprf} to
obtain the NESS pressure,
\begin{equation}
\frac{\Pi_s}{\Pi_L}=x^\gamma+\frac{\xi}{d}\frac{v^2_s(x)}{c_L^2}x^\alpha \,.
\label{solvp}
\end{equation}

\begin{figure}[b]
	\begin{subfigure}{0.49\textwidth}
		\includegraphics[width=.96\linewidth]{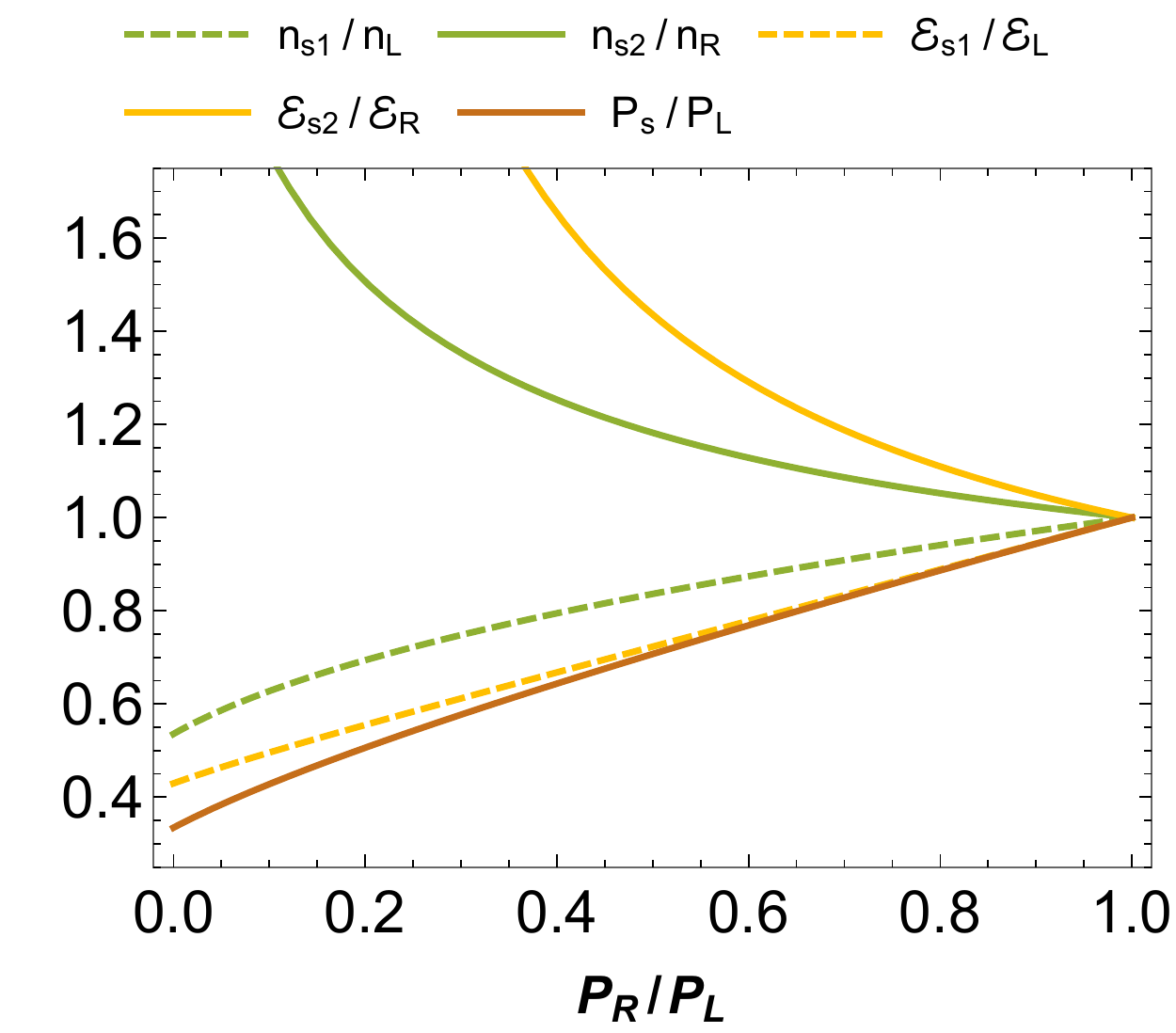}
		\label{fig:Fig 6}
	\end{subfigure}
	\begin{subfigure}{0.49\textwidth}
		\vspace{1.2cm}
		\includegraphics[width=.99\linewidth]{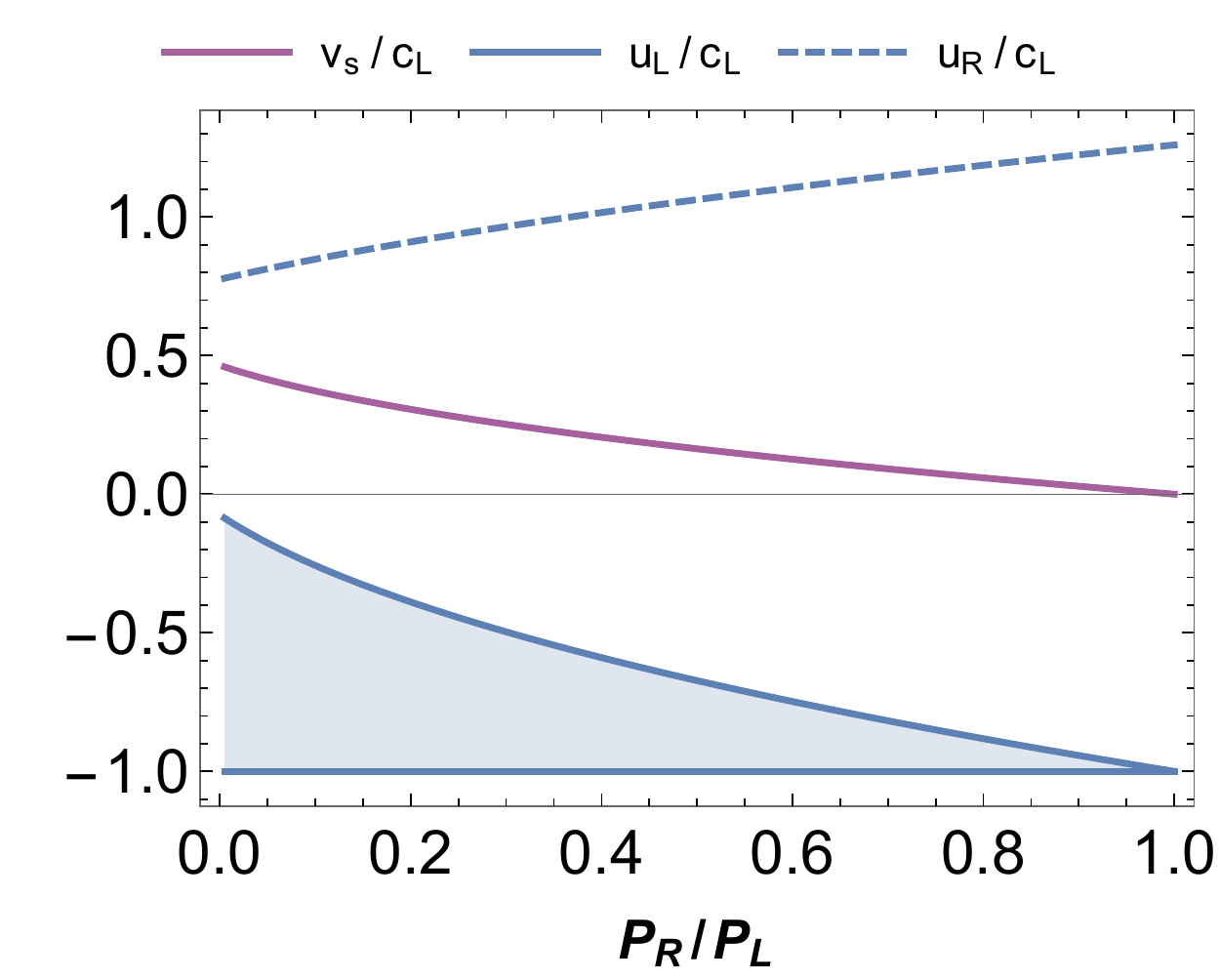}
		\label{fig:Fig 7}
	\end{subfigure}
	\caption{NESS variables for $z=3$, $d=3$ as a function of $P_R/P_L$ for fixed $n_L/n_R=2$.}
	\label{fig:Fig 6-7}
\end{figure}

\begin{figure}[t]
	\begin{subfigure}{0.49\textwidth}
		\includegraphics[width=.96\linewidth]{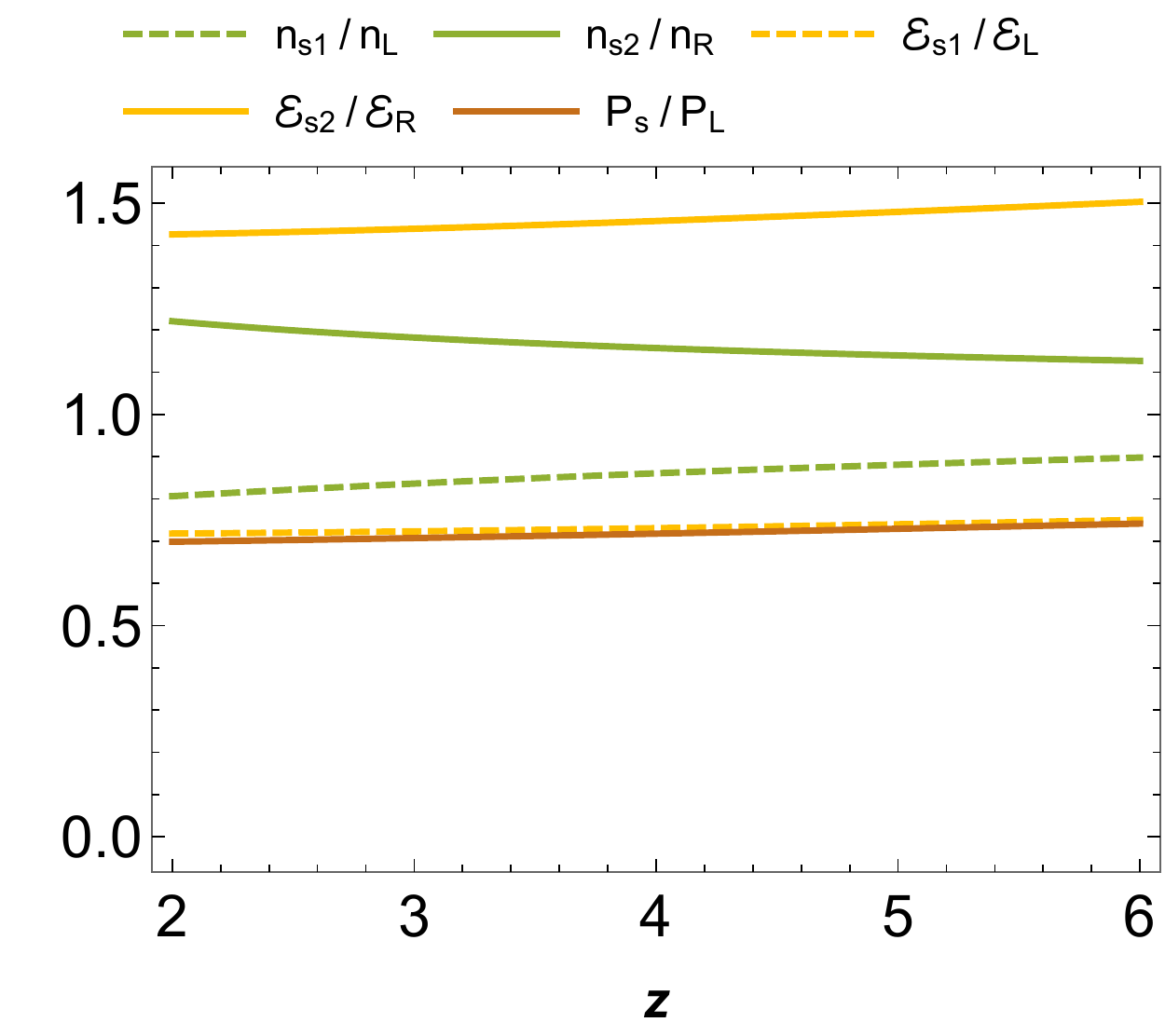}
		\label{NEPofz}
	\end{subfigure}
	\begin{subfigure}{0.49\textwidth}
		\vspace{1.2cm}
		\includegraphics[width=.99\linewidth]{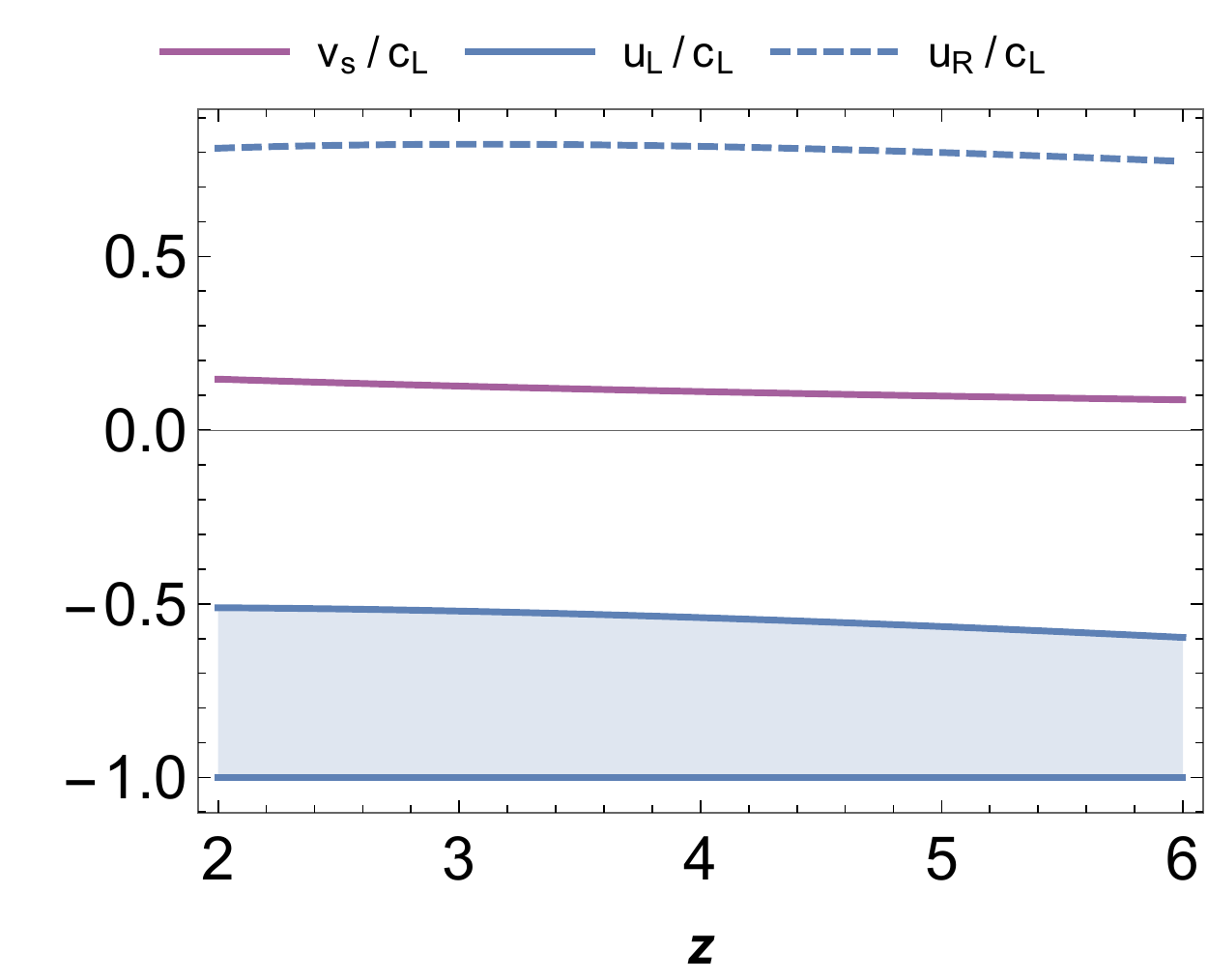}
		\label{velocitiesofz}
	\end{subfigure}
	\caption{NESS variables as a function of $z$ for $d=3$ and fixed $P_L/P_R=n_L/n_R=2$.}
	\label{NEPvelocitiesofz}
\end{figure}

The requirement that $v_s$ and $P_s$ take the same values on both sides of the contact discontinuity gives
rise to two independent relations between the variables $x$ and $y$, which is sufficient to determine their values
for given initial data for the reservoirs.\footnote{As in the $z=2$ case, the initial data only enters through the ratios 
	$P_L/P_R$ and $n_L/n_R$.} The remaining NESS variables are easily obtained once the dimensionless 
charge densities $x$ and $y$ have been solved for numerically. For instance, the NESS pressure is obtained
by inserting $y$ into the equation on the right in \eqref{yandPsol}, while the shock wave speed and the fluid speed 
in the NESS region are given by \eqref{shockspeed_z} and \eqref{nessflowspeed}, respectively. Solutions for $d=3$ 
spatial dimensions and $z=3$ are presented in Figure~\ref{fig:Fig 6-7} as a function of $P_R/P_L$. 
Figure~\ref{NEPvelocitiesofz} shows how the solution changes with $z$ for a particular choice of $P_L/P_R$
and $n_L/n_R$.

As stated above, the NESS for $z\neq 2$ cannot be recognized as a boosted thermal fluid. 
The equation of state \eqref{neweos} is incompatible with the Galilean boost transformations \eqref{estqcf}, 
which leave invariant $P$ and $n$ while shifting $\mathcal{E} \to \mathcal{E}+\frac{1}{2}n\, v^2$. 
Furthermore, the momentum density $\mathcal{P}=\rho\,v$ does not match the one obtained from a Galilean boost. 
Therefore no temperature can be associated to the solution obtained here. It is genuinely a non-thermal out-of-equilibrium 
state in a theory without boost symmetry.

\begin{figure}[t]
	\centering
	\includegraphics[width=.52\linewidth]{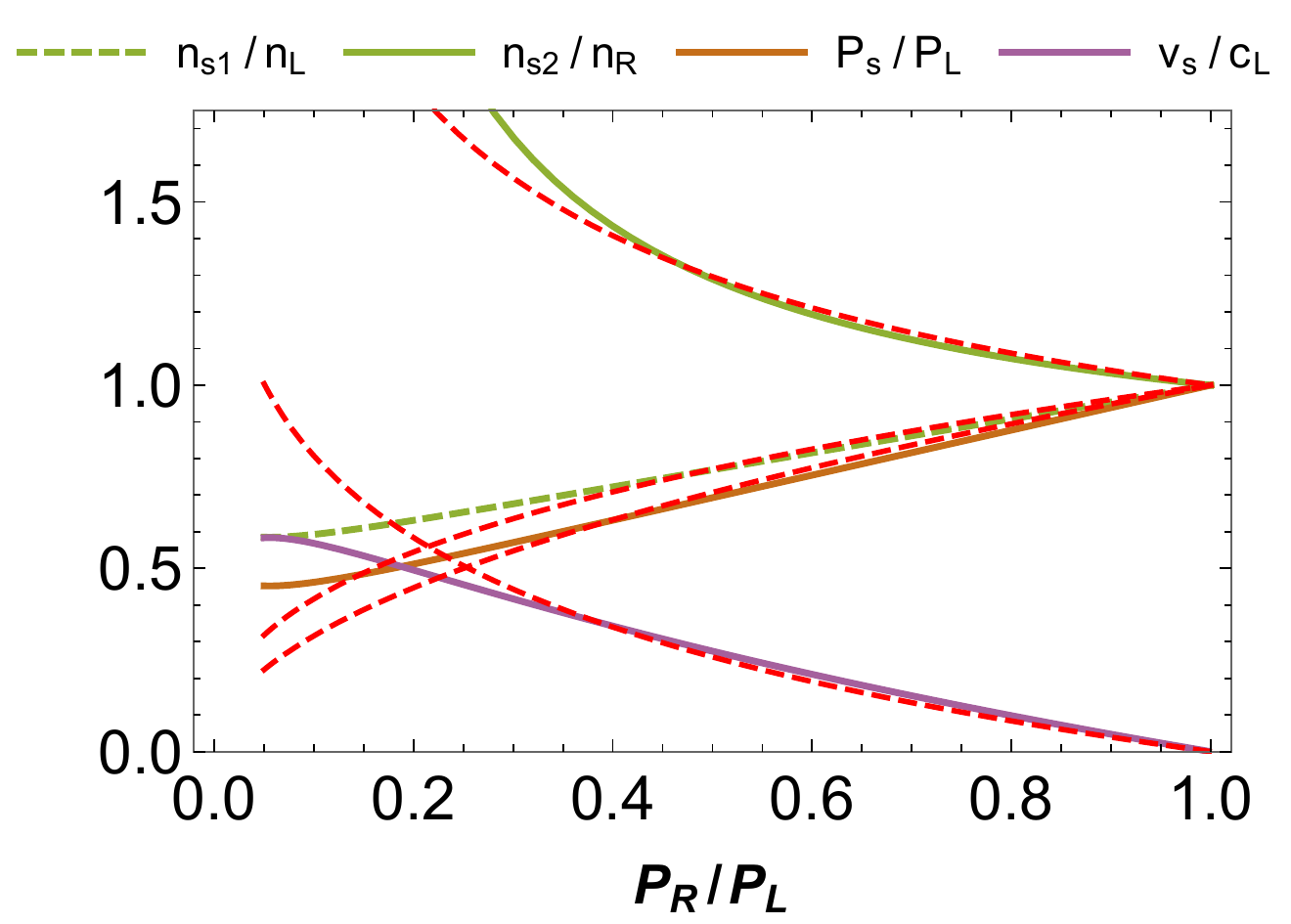}
	\caption{NESS variables for a Lifshitz fluid at $d=3$, $z=1$ and $n_L/n_R=2$ as a function of $P_R/P_L$.
		For comparison, the corresponding variables for a relativistic fluid considered in \cite{PhysRevD.94.025004,Spillane:2015daa}
		are shown by the red dashed curves. The solutions are well matched as $P_R/P_L\rightarrow 1$.}
	\label{fig:z1compare}
\end{figure}

It is interesting to compare the NESS variables we find at $z=1$ to the solution of the corresponding
Riemann problem for a relativistic fluid presented in \cite{PhysRevD.94.025004,Spillane:2015daa} in 
the limit of low flow velocity. The steady state flow is slow when $P_R/P_L$ is close to 1, {\it i.e.} when
the pressure difference between the two reservoirs is small. Figure~\ref{fig:z1compare} shows the NESS
variables $n_{s1}$, $n_{s2}$, $\Pi_s$, and $v_s$ at different values of $P_R/P_L$ for $d=3$, $z=1$,
and $n_L=2n_R$. The corresponding variables in a relativistic fluid (taken from \cite{PhysRevD.94.025004})
are indicated by red dashed curves in the figure. We see a close match for all the NESS variables as
$P_R/P_L\rightarrow 1$. 

In the relativistic case, the charge density decouples from the equations that
determine the steady state pressure and flow speed but in general this is not the case for our 
non-relativistic Lifshitz fluids. The decoupling of the charge density is, however, recovered in the limit of 
small pressure difference in the $z=1$ Lifshitz case. To see this, one carries out an expansion in powers
of small $\Delta = \Pi_L-1$ in \eqref{solvx} and \eqref{solvp} that determine $\Pi_s$ and $v_s$ at $z=1$
and observes that $\eta=n_L/n_R$ indeed decouples from the equations to leading order in $\Delta$. 
For large values of $\Delta$ the steady state flow speed is no longer small and there is no reason to 
expect a match between a relativistic fluid and a $z=1$ Lifshitz fluid. 

\section{Discussion}
\label{conclusion}

The above study of the Riemann problem for Lifshitz fluids had a twofold purpose. On the one hand, 
it extends to a non-relativistic setting some recent work on the out-of-equilibrium flow of relativistic quantum 
critical fluids \cite{Bernard:2012je,Bhaseen:2013ypa,PhysRevD.94.025004,Spillane:2015daa,Pourhasan2016}, 
and, on the other hand, it provides an application to a concrete physical setup of a recently developed general 
formalism for perfect fluids without boost symmetry \cite{10.21468/SciPostPhys.5.1.003}. 

We have established that a non-equilibrium steady state, of the type seen previously 
in a relativistic scale invariant fluid, will also develop in a non-relativistic critical fluid when 
two reservoirs are brought into contact across a hypersurface. Consistent with the Lax entropy 
conditions, the non-relativistic NESS is bounded on one side by an outgoing shock wave and 
on the other side by a rarefaction wave propagating in the opposite direction. Inside the NESS
there is a contact discontinuity where the charge density jumps but the pressure stays unchanged.

In the special case of a $z=2$ Lifshitz fluid the NESS is a Galilean boost of a thermal 
equilibrium state, in direct analogy with the Lorentz boosted thermal state seen in the 
corresponding relativistic problem. Using a simple scaling ansatz for the kinetic mass density 
of a Lifshitz fluid at generic $z$, we found that the fluid variables in the central region can be 
solved for and a NESS forms in this case as well, but the solution is genuinely non-thermal.

There are several future directions to be explored. In this study, we have concentrated on 
perfect fluids without impurities or lattice effects which break translational invariance. 
Proceeding along the lines of \cite{PhysRevD.94.025004}, where this has been done for 
a conformal fluid, one could allow for diffusion and momentum relaxation in the 
hydrodynamics equations, to obtain the time scale up to which the non-relativistic 
NESS persists. 

Another interesting direction is to analyse a dual gravitational description of non-equilibrium
steady states of Lifshitz fluids. In this context, it would especially be interesting to identify a 
gravitational dual of a $z \neq 2$ Lifshitz fluid flow without boost symmetry. 

\acknowledgments
We thank Jelle Hartong, Niels Obers, Napat Poovuttikul, and Watse Sybesma for useful discussions. 
This work was supported in part by the Icelandic Research Fund grant 195970-051 and the University of Iceland Research Fund.

\bibliographystyle{JHEP}

\end{document}